\documentclass[twocolumn,prb,aps,floatfix,longbibliography,superscriptaddress]{revtex4-2}

\usepackage{color}
\usepackage{graphicx}
\usepackage{siunitx}
\usepackage[color=green!60,textsize=small]{todonotes}
\usepackage{physics}
\usepackage{amsthm}
\usepackage{amsmath}
\usepackage{amssymb}
\usepackage{enumerate}
\usepackage{placeins}
\usepackage{mathtools}
\usepackage{booktabs}
\usepackage{dsfont}
\usepackage{yfonts}
\usepackage{calligra}
\usepackage{hyperref}
\usepackage{svrsymbols} 
\usepackage{tikz}

\newcommand{\cthird}{%
\begin{tikzpicture}[scale=0.25,baseline=+0.25ex]%
\draw (0,1) -- (0.57735,0) -- (-0.57735,0) -- cycle; 
\draw[black,fill=black] (-0.57735,0) circle (1.0ex);
\end{tikzpicture}} 

\newcommand{\cfull}{
\begin{tikzpicture}[scale=0.25,baseline=+0.25ex]
\draw (0,1) -- (0.57735,0) -- (-0.57735,0) -- cycle; 
\draw[black,fill=black] (-0.57735,0) circle (1.00ex);
\draw[black,fill=black] (0.57735,0) circle (1.00ex);
\draw[black,fill=black] (0,1) circle (1.00ex);
\end{tikzpicture}} 

\DeclareMathAlphabet{\mathcalligra}{T1}{calligra}{m}{n}
\DeclareFontShape{T1}{calligra}{m}{n}{<->s*[2.2]callig15}{}

\renewcommand{\vec}[1]{\boldsymbol{#1}}
\newcommand{\Eqref}[1]{Eq.~\eqref{#1}}

\usepackage{array}
\usepackage{hhline}
\usepackage{latexsym}
\usepackage{bm}
\usepackage{bbm}
\usepackage{cancel}

\newcommand{\be}{\begin{equation}}
\newcommand{\ee}{\end{equation}}
\newcommand{\bsube}{\begin{subequations}}
\newcommand{\esube}{\end{subequations}}
\newcommand{\ba}{\begin{array}}
\newcommand{\ea}{\end{array}}

\newcommand{\bea}{\begin{eqnarray}}
\newcommand{\eea}{\end{eqnarray}}

\newcommand{\Vext}{\mathcal{V}_{\rm He-\graphene}}
\newcommand{\Vhehe}{\mathcal{V}_{\rm He-He}}

\newcommand{\vrg}{\!\mathbf{\mathcalligra{r}}}
\newcommand{\rg}{\mathcalligra{r}}

\newcommand{\tilVHeG}{\widetilde{\mathcal{V}}_{\rm He-\graphene}}

\AtBeginDocument{%
\heavyrulewidth=.08em
\lightrulewidth=.05em
\cmidrulewidth=.03em
\belowrulesep=.65ex
\belowbottomsep=0pt
\aboverulesep=.4ex
\abovetopsep=0pt
\cmidrulesep=\doublerulesep
\cmidrulekern=.5em
\defaultaddspace=.5em
}

\begin{document}

\title{Two-Dimensional Bose-Hubbard Model for Helium on Graphene}

\author{Jiangyong Yu}
\affiliation{Department of Physics, University of Vermont, Burlington, VT 05405, USA}

\author{Ethan Lauricella}
\affiliation{Department of Physics, University of Vermont, Burlington, VT 05405, USA}

\author{Mohamed Elsayed}
\affiliation{Department of Physics, University of Vermont, Burlington, VT 05405, USA}

\author{Kenneth Shepherd Jr.}
\affiliation{Department of Physics, University of Vermont, Burlington, VT 05405, USA}

\author{Nathan S. Nichols} 
\affiliation{Department of Physics, University of  Vermont, Burlington, VT 05405,
USA}
\affiliation{Materials Science Program, University of Vermont, Burlington, VT 05405,USA}

\author{Todd Lombardi}
\affiliation{Department of Physics and Astronomy, University of Missouri, Columbia, MO 65211, USA}

\author{Sang Wook Kim}
\affiliation{Department of Physics, University of Vermont, Burlington, VT 05405, USA}

\author{Carlos Wexler} 
\affiliation{Department of Physics and Astronomy, University of Missouri, 
Columbia, MO 65211, USA}

\author{Juan M.~Vanegas}
\affiliation{Department of Physics, University of Vermont, Burlington, VT 05405, USA}

\author{Taras Lakoba} 
\affiliation{Department of Mathematics \& Statistics, University of  Vermont, Burlington, VT 05405, USA}

\author{Valeri N.~Kotov}
\affiliation{Department of Physics, University of  Vermont, Burlington, VT 05405, USA}

\author{Adrian Del Maestro}
\affiliation{Department of Physics and Astronomy, University of Tennessee, Knoxville, TN 37996, USA}
\affiliation{Min H.~Kao Department of Electrical Engineering and Computer Science, University of Tennessee, Knoxville, TN 37996, USA}
\affiliation{Department of Physics, University of Vermont, Burlington, VT 05405, USA}

\begin{abstract}
An exciting development in the field of correlated systems is the possibility of realizing two-dimensional (2D) phases of quantum matter. For a systems of bosons, an example of strong correlations
manifesting themselves in  a 2D environment is provided by helium adsorbed on graphene.
We construct the effective Bose-Hubbard model for this system which involves hard-core bosons $(U\approx\infty)$, repulsive nearest-neighbor $(V>0)$ and small attractive $(V'<0)$ next-nearest neighbor interactions. The mapping onto the Bose-Hubbard model is accomplished by a variety of many-body techniques which take into account the strong He-He correlations on the scale of the graphene lattice spacing. Unlike the case of dilute ultracold atoms where interactions are effectively point-like, the detailed microscopic form of the short range electrostatic and long range dispersion interactions in the helium-graphene system are crucial for the emergent Bose-Hubbard description.  The result places the ground state of the first layer of $^4$He adsorbed on graphene deep in the commensurate solid phase with $1/3$ of the sites on the dual triangular lattice occupied. Because the parameters of the effective Bose-Hubbard model are very sensitive to the exact lattice structure, this opens up an avenue to tune quantum phase transitions in this solid-state system.
\end{abstract}

\maketitle

\section{Introduction}

\subsection{Helium on Two-Dimensional Materials: A Many-Body Paradigm}
\label{subsec:heliumgeneral}

The problem of $^4$He atoms deposited on solid substrates has been identified
for many decades as a bosonic many-body problem that could exhibit a rich phase
diagram including the possibility of dimensional crossover \cite{Bretz:1971jo,Bretz:1973ky, Dash:1979bk, Greywall:1993kn, Dash:1994cp,Gasparini:2008ka, Reatto:2013hz, Makiuchi:2018km, Saunders:2018em,Saunders:2019em}. Graphite was first recognized as an ideal
two-dimensional substrate due to its exceptional homogeneity,
\cite{Thomy:1969ti} and extensive experimental \cite{Bretz:1973ky, Crowell:1996kn,Nyki:1997ss} and theoretical studies \cite{Whitlock:1998gb, Pierce:1999cn,Corboz2008cb} have demonstrated that under the right circumstances a superfluid He film can develop on the graphite surface. Because $^4$He atoms are neutral, the many-body interactions that determine the behavior of this system are the van der Waals (VDW) interactions between He atom pairs and between He and graphite. Since VDW interactions are typically fairly weak, but long range, the possibility of superfluidity, and at which density (and film coverage) it can exist depends on the interplay between the two-body He--He interactions and the interaction of He with the substrate (in this case carbon) atoms.

Since the discovery of the two-dimensional (2D) version of graphite, namely graphene \cite{Antonio}, the problem of He--substrate interactions has been revisited with great enthusiasm \cite{Nichols:2016hd,Gordillo:2009jb, Reatto:2013hz,Happacher:2013ht}.  As graphene is a purely 2D system,  the VDW adsorption potential that tends to localize helium-4 atoms is 10\% weaker (compared to graphite which is a bulk material) and therefore there is the exotic possibility of purely 2D ${^4}$He superfluidity (atomic width film). 
While graphite's properties are set by its bulk structure, graphene's 2D lattice and (related) electronic structure  can be manipulated in a variety of ways. This is the reason why graphene and 2D materials more generally have become an attractive area of theoretical and applied electronics research \cite{vdwhetero}. For example, doping (addition of electrons or holes into the layer) can be easily done, or the hexagonal structure  can be distorted, or hydrogenation agents can be introduced (making graphene effectively an insulator) \cite{Antonio,Kotov}. All of these affect the graphene lattice and electronic state and, by extension, the VDW  potential between He and graphene \cite{Nichols:2016hd}. Finally, graphene's dielectric environment can be easily changed. For example, putting graphene on different dielectric substrates immediately affects (screens) the electronic charge resulting in a modified strength of the VDW force.
 
For all of the above reasons the problem of $^4$He on graphene, and its extensions, has become a pressing problem due to its potential to produce purely 2D collective bosonic states. The first question to answer is the behavior of helium-4 on pristine graphene in vacuum. So far, theoretical studies \cite{Gordillo:2009jb, Bruch:2010fq, Gordillo:2011jb, VranjesMarkic:2013br, Happacher:2013ht} have concluded that the first adsorbed He layer on graphene forms an insulating state where He atoms occupy 1/3 of all graphene hexagon centers (energetically preferred location), in a triangular lattice pattern, the so-called C1/3 commensurate solid.
%
\begin{figure}[t]
\begin{center}
    \includegraphics[width=\columnwidth]{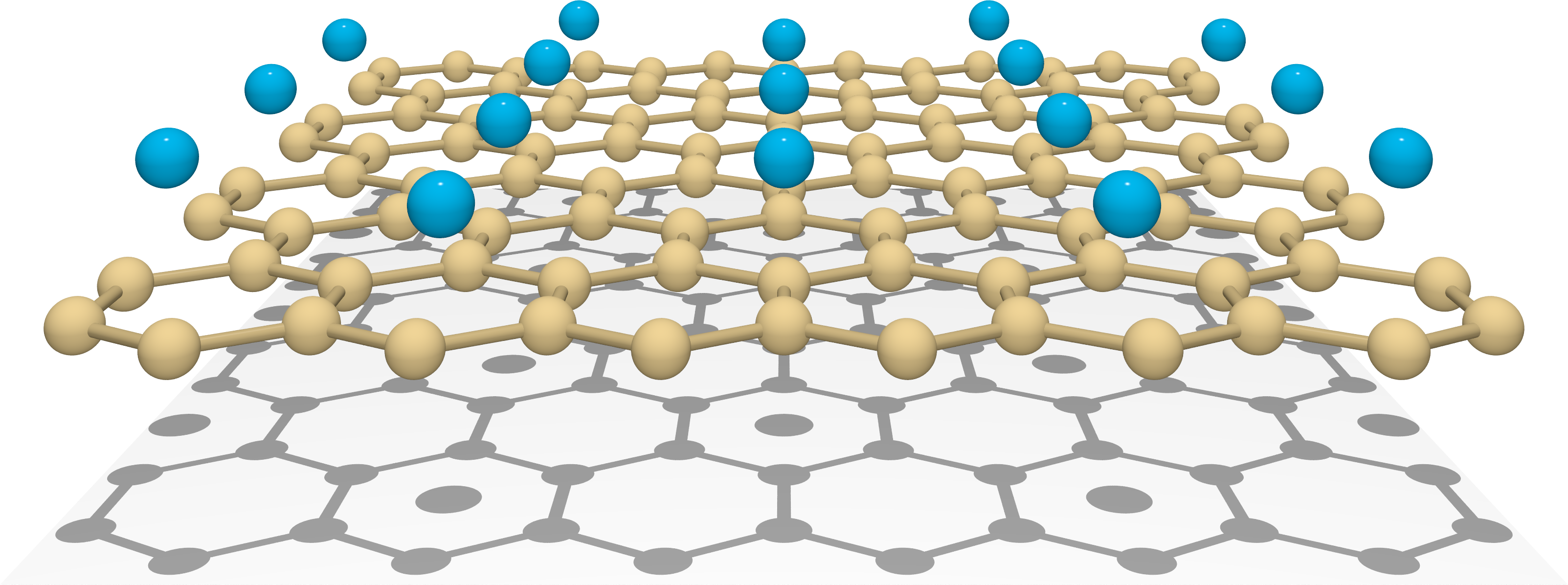}
\end{center}
\caption{A schematic of the $\sqrt{3} \times \sqrt{3}$ commensurate solid phase for $^4$He atoms (blue) adsorbed on graphene (gold) showing $1/3$ of the $N_\graphene = 72$ sites filled where the He atoms are localized on the sites of a triangular lattice (see shadow). The size of the indicated region of graphene is 
$L_x \times L_y \simeq \SI{22}{\angstrom} \times \SI{17}{\angstrom}$.}
\label{fig:cell}
\end{figure}
%
There is still some limited controversy on the possible existence of a competing classical or quantum liquid at zero temperature \cite{Gordillo:2011jb} based on the exact form of the helium--graphene potential utilized in simulations, but energy differences are on the order of statistical uncertainties. 
 This pinning of He atoms in this insulating state (depicted schematically in Fig.~\ref{fig:cell}) is due to a combination of the He--graphene attractive VDW potential and He--He repulsion, as we discuss below. The second He layer can become superfluid \cite{Kwon:2012ie, Gordillo:2012fl,Happacher:2013ht}, as it is farther away from the attractive graphene potential, even though studies show that a number of other states, including incommensurate solid phases, are very close to it in energy.  Overall, the emergence of superfluidity turns out to be a very complex many-body problem due to a fine balance between fairly weak VDW forces.
 
The aim of this work is to conclusively develop an effective 2D Bose--Hubbard (BH) model for the first layer of ${^4}$He on graphene. The reasons why such a model is highly desirable are as follows.  (1) The results mentioned above about the existence of the 1/3 insulating state  are obtained by different zero and finite temperature quantum Monte Carlo (QMC) techniques. In fact, we will complement those with our own version of ground state continuum QMC. However, to gain intuition about the stability of the C1/3 phase and its proximate phases, it is advantageous to develop an effective lattice Bose--Hubbard model where the most important interactions are identified. Of course, the phases predicted by the effective BH model must agree with the QMC results. We will see that this is indeed the case.  (2) It is clear from the outset that the resulting BH model is highly non-trivial to develop, compared, for example, with BH models used in cold atom physics (where optical lattice potentials are the equivalent of the graphene potential here). The reason is that in cold atom physics the atom density is very low (a billion times lower), while in our case of He on graphene the coverage is high, and atoms are separated from each other on the scale of the graphene lattice (several Angstroms), which is smaller than the range of the VDW potential. Thus, while interaction effects in cold atom physics are generally easy to incorporate by assuming $s$-wave scattering between atoms \cite{Jaksch:1998,Jaksch:2005ym, Bloch:2008, Jaksch:2013}, this is not the case in our solid-state context where there is a finite range over which interactions are important.  It is not \emph{a priori} clear that a consistent 2D effective BH model description even exists since the QMC techniques previously mentioned are fully 3D. Thus a careful comparison between ``2D restricted" QMC and several other techniques has to be made. 
(3) Finally, armed with such an effective 2D BH model, one can use it as a first step in the  analysis of a  variety of other systems, including situations where graphene's properties are modified (as previously described), or generalizing to other 2D materials. 

The overall ``raison d'\^{e}tre" of a reliable Bose-Hubbard description is   that it allows studies of strongly correlated phases, such as supersolids, correlated insulators and superfluids,  as well as the quantum phase transitions between them. The Bose-Hubbard model is an effective low-energy Hamiltonian which in itself represents a lattice many-body problem; however its properties and phases are more amenable to theoretical analysis than the original many-body description, especially when the types and values of the Bose-Hubbard parameters can be reliably extracted from the original model. Consequently the  Bose-Hubbard  description can  also be used as a powerful tool to 
 ``predict" the existence of quantum phases with specific properties on the basis of the relationship between the Bose-Hubbard parameters and the various interactions in the full microscopic model.

\subsection{Approach and Summary of Main Results}
\label{subsec:approach}

The main result of this work is that the behavior of the first layer of $^4$He atoms adsorbed on graphene can be captured via a single-band ``hard-core'' Bose--Hubbard model with strong (effectively infinite) on-site Hubbard repulsion ($U\approx \infty$).  This is in contrast to previous studies in the second and even higher adsorbed layers, that considered a phenomelogical soft-core BH model \cite{Zimmerli:1988ii}.  

\begin{figure}[t]
\begin{center}
    \includegraphics[width=\columnwidth]{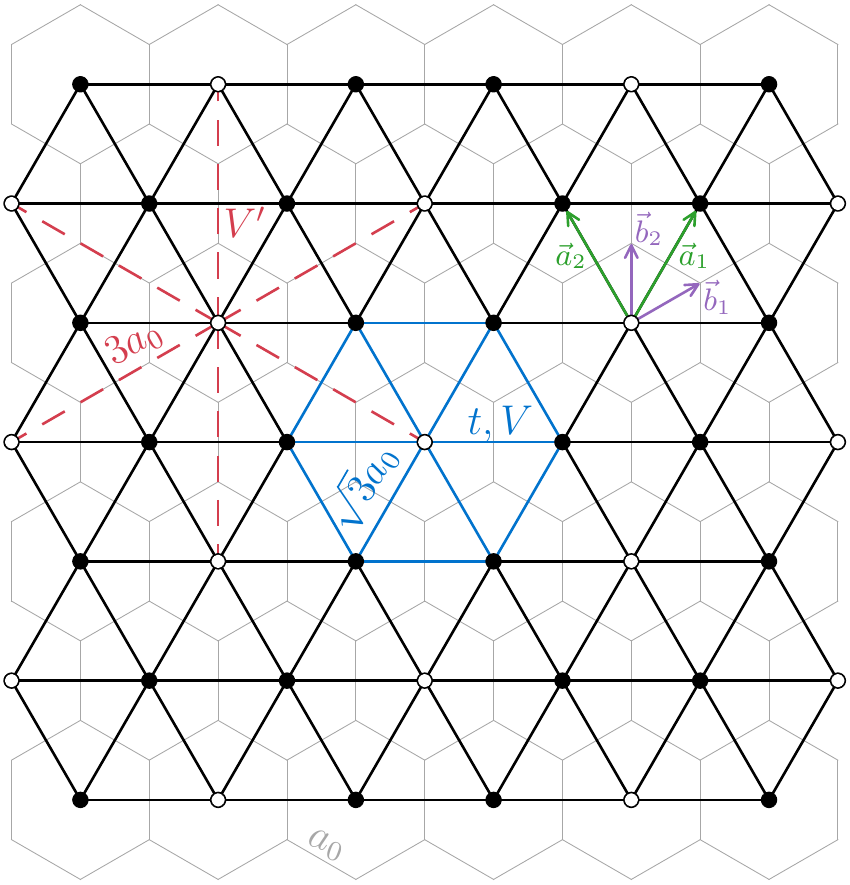}
\end{center}
\caption{The triangular lattice defined by hexagon centers of the graphene
lattice (shown in grey) with the lattice and basis vectors in Eq.~\eqref{eq:grapheneBasis} shown in the upper right.  The nearest neighbor hopping $t$ and interaction $V$ are experienced between sites separated by $\sqrt{3}a_0$ (blue) where $a_0 \simeq \SI{1.42}{\angstrom}$ is the length of a hexagon side.  The dashed red lines indicate the next-nearest neighbor interaction $V^\prime$ between sites separated by $3a_0$ which form a triangular superlattice at $1/3$ filling (open circles) which corresponds to the adsorbed C1/3 solid depicted in Fig.~\ref{fig:cell}. Lattice vectors $\vb*{a}_1$ and $\vb*{a}_2$ are shared between the graphene and triangular lattice.}  
\label{fig:triangular}
\end{figure}

For the first layer considered here, we find that the resulting low energy Hamiltonian has the form:
\begin{align}
    H_{BH} & = -t \sum_{\langle i,j\rangle} (b_{i}^\dagger b_j^{\phantom \dagger} + h.c.) + V  \sum_{\langle i,j\rangle} n_i n_j \nonumber \\
           & \quad+ V^\prime \sum_{\langle \langle i,j \rangle \rangle} n_i n_j + \dots,
           \label{eq:BHHamiltonian}
\end{align}
where $t$ is the hopping strength, $b_i^{{\dagger}} (b_i^{\phantom{\dagger}})$ creates (destroys) a bosonic $^4$He atom on site $i$ with $[b_i^{\phantom\dagger},b_j^{{\dagger}}]=\delta_{i,j}$, $V$ is the nearest neighbor interaction, and $V^\prime$ is the next-nearest neighbor interaction. The ellipsis indicates higher order interactions that are neglected here.  The sites $i,j$ correspond to the vertices of the triangular lattice formed by the centers of graphene's hexagons as seen in Fig.~\ref{fig:triangular}.
%
%
%
We find by a variety of methods that: $t \sim \SI{1}{K}$, $V\sim \SI{50}{\kelvin}$ and $V^\prime \sim \SI{-2}{\kelvin}$. A detailed comparison of the different methods we employ, and the assumptions inherent in their use, forms the bulk of this study.

We start with the one particle properties, and  in order to compute the hopping $t$ we employ maximally localized Wannier function. First, the VDW potential due to graphene, acting on a single He atom is calculated by techniques described in our previous work \cite{Nichols:2016hd}. Then the one-particle Schrodinger equation in the external VDW potential is solved numerically and the Wannier functions are constructed. The electronic dispersion follows the symmetry of the triangular lattice and  the effective hopping $t$ from this analysis is $\sim \SI{1}{\kelvin}$.

We also estimate $t$ by independently computing the effective 2D adsorption potential experienced by a $^4$He atom above the graphene sheet using path integral ground state quantum Monte Carlo (QMC) and two types of {ab initio} methods: Density Functional Theory (DFT) \cite{QE09,QE17} improved by including VDW energies in the appropriate DFT functional \cite{D4VdW17,D4VdW19,D4VdW20}, and 2nd order M\o{}ller--Plesset (MP2) \cite{Moller:1934hc,Cramer,Bartlett}. In all cases, the non-interacting band structure resulting from the periodic adsorption potential is determined, where the hopping can then be extracted from the bandwidth or overlap integrals.  The results are all consistent with the simple Wannier theory and provide a powerful confirmation of the accuracy of the extracted value of $t$. 

Next, we turn to the He--He interaction induced terms $(U,V,V')$ in Eq.~\eqref{eq:BHHamiltonian}.  The consistent incorporation of interactions proves rather formidable and introduces unique technical challenges, not usually encountered when building and estimating parameters in an effective Bose--Hubbard model applicable to ultra-cold bosonic lattice gases \cite{Jaksch:1998,Jaksch:2005ym,Becker:2010jf,Jaksch:2013,Ibaez:2013tb}.  The He--He potential itself between two isolated atoms in  vacuum is well understood and can be very accurately parametrized, as a result of decades of study \cite{Aziz:1979hs, Aziz:1987ia, Aziz:1995kc, Przybytek:2010ol,Cencek:2012iz}.  The first observation is that $^4$He is special because the $s$-wave scattering length $a_s \sim \SI{100}{\angstrom}$ \cite{Tang:1995yw,Grisenti:2000uf} is ``by chance" (i.e. without any fine-tuning) very large, of the order $>10$ times larger than the range of the potential (the effective VDW length). 
For very dilute lattice systems of trappable heavy atoms,  the average atomic separation $\ell \gg a_s$, and interactions in an effective BH description can be computed by convolving highly localized spatial atomic wavefunctions with narrow $\delta$-function interactions with a strength proportional to $a_s$.

For the adsorption geometry considered here -- helium atoms confined to move on a triangular lattice with spacing $a \sim \SI{2}{\angstrom}$ due to the proximate graphene structure -- we are in the opposite limit where the calculation of effective interaction parameters is very sensitive to the short-range part of the He--He potential.  At small scales, below $\sim \SI{2}{\angstrom}$, this potential rises rapidly to a very large strength $\sim \SI{E6}{\kelvin}$ yielding a hard-core description with $U\approx\infty$ and promoting the nearest neighbor interaction $V$ to play a dominant role. In addition, the fact that He--He is very strong at the lattice scale suggests that a self-consistent formulation has to be employed for the calculation of $V$, which operates on this scale.  Calculating $V$ by using the ``bare" Wannier functions, i.e.\ the single-particle localized wavefunction in the field of graphene, produces an unphysically large value $V\sim \SI{E3}{\kelvin}$.  This problem suggests a strategy where a self-consistent  adjustment of the Wannier functions to accommodate the strong repulsion is employed, for example in the spirit of the Jastrow factor commonly introduced in such situations \cite{McMillan:1965ue,Whitlock:1979pf,Lutsyshyn:2015uo}.  We have determined that instead of working with Jastrow factors, it is more convenient to use the self-consistent Hartree--Fock equations \cite{Kaxiras:2019,PethickSmith:2008}.  These are expected to provide  a very accurate description of two-body interactions due to the strongly localized nature of the Wannier functions around a given site.  We find that the Hartree--Fock equations converge to the same result (``fixed point") which is independent of the details of the potential at ultra-small distances, producing $V\sim \SI{70}{\kelvin}$ (see \S{}\ref{subsec:HF}).

The value of $V$ can also be calculated within three additional and complementary approaches.  The continuum QMC method mentioned previously, provides a very accurate estimate for the adsorbed $^4$He wavefunctions and the total interaction energy at unit filling that can be converted into an effective $V\sim 50$ K (see \S{}\ref{subsec:qmc}). This is in satisfactory agreement with the Hartree--Fock method.  Van der Waals corrected DFT provides a third, independent check of the above results which yields $V \sim \SI{20}{\kelvin}$ (\S{}\ref{subsec:DFT}) and {ab initio} 2nd order M\o{}ller--Plesset perturbation theory for two adsorbed He atoms on a variety of aromatic carbons (up to circumcoronene) yields $V \sim \SI{50}{\kelvin}$. 

The combination of all the aforementioned techniques (each subject to very different approximations) leads to an effective Bose--Hubbard model (\Eqref{eq:BHHamiltonian}) with parameters summarized in Table~\ref{tab:BHresults}. All energies are reported in \si{\kelvin}, the natural scale in the adsorption system under consideration. 
\begin{table}
    \renewcommand{\arraystretch}{1.5}
    \setlength\tabcolsep{13pt}
    \begin{tabular}{@{}lllll@{}} 
        \toprule
        Method & $t \, (\si{\kelvin})$ & $V \,(\si{\kelvin})$ & $V^\prime
        \, (\si{\kelvin})$ & ${t}/{V}$ \\ 
        \midrule
        Wannier & 1.45 & 7540 & 638 & 0.0002 \\
        HF & 1.45 & 69.7 & -2.08 & 0.021 \\
        QMC & 1.38 & 54.3(1) & -2.76(2) & 0.025 \\
        DFT  & 1.10 & 21.4 & -1.36 & 0.051 \\
        MP2  & 0.59 & 51.5 & -1.97 & 0.011 \\
        \bottomrule
    \end{tabular}
    \caption{\label{tab:BHresults} The hopping parameter $t$, nearest and next nearest-neighbor interaction $V$ and $V'$, and the ratio of $t/V$ of the effective Bose--Hubbard model defined in \Eqref{eq:BHHamiltonian} as calculated by the five different methods: Wannier functions, Hartree--Fock (HF), quantum Monte Carlo (QMC), Density Functional Theory (DFT), and M\o{}ller--Plesset perturbation theory (MP2). In all cases, $t$ is calculated via the band structure of a single helium atom subject to a periodic two-dimensional adsorption potential $\Vext$.  Note that $t$ is the same for Wannier and Hartree--Fock as they use the same empirical potential.}
\end{table}
With the exception of the simple Wannier theory (which as discussed above does not properly take into account the effects of interactions on the lattice scale), all results are in good agreement, allowing us to definitively place helium on graphene within the context of the extended hard-core Bose--Hubbard model on the triangular lattice. 

\subsection{Implications for the Quantum Phase Diagram}
\label{subsec:summary}

\begin{figure}[h]
\begin{center}
    \includegraphics[width=\columnwidth]{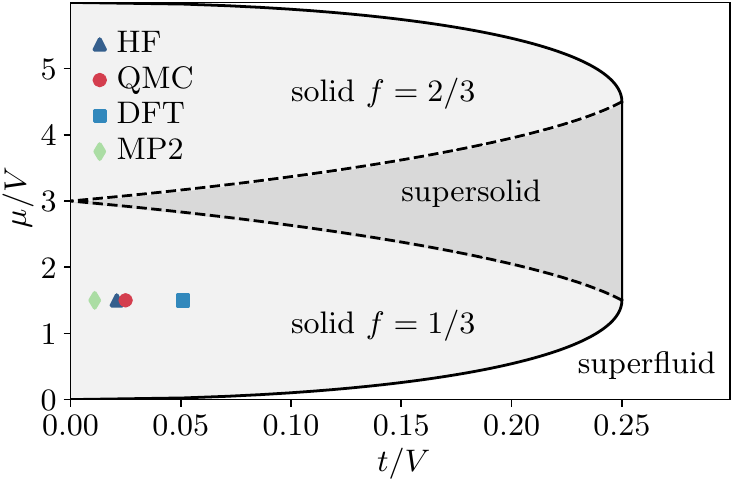}
\end{center}
\caption{The mean-field phase diagram for hard-core bosons on the triangular lattice with nearest-neighbor interactions $V$ and hopping $t$ with density controlled by the chemical potential $\mu$.  Identified phases include commensurate solids (at fillings $f=1/3,2/3$), superfluid, and supersolid (a superfluid that breaks triangular lattice symmetries). Solid lines indicate discontinuous (first order) transitions, while continuous (second order) transitions occur across dashed lines.  The data points in the lower left-hand corner represent the major results of this paper, indicating that the ground state of a single layer of $^4$He on graphene resides deep in the commensurate solid phase at $1/3$ filling. For these data points, the chemical potential has been chosen such that $\mu/V$ has the same value as the tip of the first lobe.}
\label{fig:PhaseDiagram}
\end{figure}

The phase diagram of \Eqref{eq:BHHamiltonian} in the limit of infinite $U$ and considering only nearest neighbor interactions ($t-V$ model) can be analyzed within the mean-field theory \cite{phase-diag,Murthy-phase-diag}, as shown in Fig.~\ref{fig:PhaseDiagram}. This result is known to be in qualitative agreement with lattice quantum Monte Carlo for hard-core bosons with extended interactions \cite{Wessel:2005ik, Gan:2007zd, Zhang:2011iz}.
For small values of the chemical potential (low filling fraction) three phases
are identified: the C1/3 phase discussed previously, a supersolid phase, and
uniform superfluid phase. The phase boundary between the solid $f=1/3$ state
and the uniform superfluid state is evaluated by comparing the ground state energies of
the respective configurations.  The $f=1/3$ state is characterized by density
wave order as shown in Figures~\ref{fig:cell} and \ref{fig:triangular} (one atom per triangular unit cell), while the superfluid breaks no translational symmetries, but exhibits a finite (uniform) superfluid density.

Our parameter ratio $t/V \sim 1/50$ (Table~\ref{tab:BHresults}), places $^4$He on pristine graphene firmly in the C1/3 phase, (as shown by the symbols) consistent with previous simulations of the full three dimensional system \cite{Happacher:2013ht}.

\subsection{Paper Outline}
\label{subsec:outline}
In the remainder of the paper, we provide a discussion of the microscopic models we employ to characterize a three-dimensional system of helium atoms interacting with a two-dimensional graphene membrane.  We then discuss in what context or limits this system can be understood within an effective 2D theory.  Working within the 2D limit, we provide details of the approaches briefly discussed in the introduction to estimate the parameters of a hard-core extended Bose--Hubbard model. This includes studying the band structure of a single $^4$He atom adsorbed on graphene, and determining wavefunctions for the many-particle system at different levels of sophistication.  Finally, we conclude by comparing all our results and describe the exciting future directions this work opens up for studying hard-core Bose--Hubbard models in a solid state setting.

All data and code needed to generate the results in this paper are available online \cite{repo}.

\section{Model: Helium on Graphene}

We consider a system of $^4$He atoms in proximity to a graphene substrate
frozen at $z=0$ with lattice ($\vec{a}$) and basis ($\vec{b}$) vectors:
\begin{equation}
\begin{aligned}
    \vec{a}_1 &= \frac{a_0}{2}\qty(\sqrt{3},3), & \vec{b}_1 &= \frac{a_0}{2}\qty(\sqrt{3},1) \\
    \vec{a}_2 &= \frac{a_0}{2}\qty(-\sqrt{3},3), & \vec{b}_2 &= a_0\qty(0,1) 
\label{eq:grapheneBasis}
\end{aligned}
\end{equation}
where $a_0 \simeq \SI{1.42}{\angstrom}$ is the carbon--carbon distance as depicted in Figs.~\ref{fig:cell} and \ref{fig:triangular}.   In this paper we consider two classes of methods distinguished by how interactions are handled. (A) In tight binding, Hartree--Fock, and quantum Monte Carlo calculations we employ empirical interaction potentials, while (B) 
van der Waals corrected density functional theory \cite{QE09,QE17} and M\o{}ller--Plesset \cite{Moller:1934hc} perturbation theory utilizes an {ab initio} estimate for the interaction energy within the Born--Oppenheimer approximation.  The combination of these two classes of methods ensures a broader regime of applicability and improved confidence in our final results for the mapping of the microscopic system to an effective extended Bose--Hubbard model.

\subsection{Empirical}
\label{subsec:empirical}
\begin{figure}[t]
\begin{center}
    \includegraphics[width=\columnwidth]{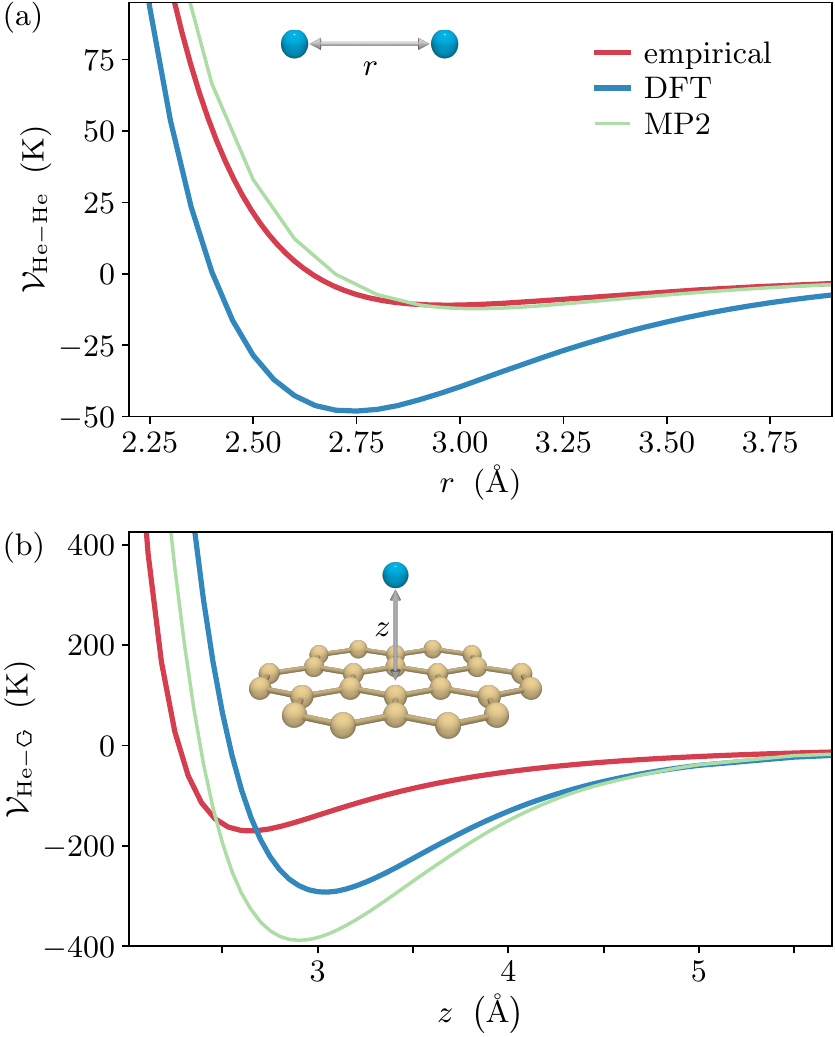}
\end{center}
\caption{The interaction (a) and adsorption (b) potentials in \Eqref{eq:Ham}.  The three curves in both panels indicate three different approaches to the potentials.   Empirical indicates the formulas discussed in the text where $\Vhehe$ is taken from Ref.~\cite{Przybytek:2010ol} while $\Vext$ is determined from \Eqref{eq:HeGraphene} at $\vrg = (0,0)$.   The potentials labelled DFT and MP2 are extracted from the minimal energy surfaces determined by those {ab initio} methods as described in \S~\ref{subsec:DFT} and \ref{subsec:MP2}.}
\label{fig:VHeGraphene}
\end{figure}

A system of $N$ $^4$He atoms of mass $m$ interacting with the graphene membrane
can be described in first quantization via the Hamiltonian:
\begin{equation}
    H = -\frac{\hbar^2}{2m} \sum_{i=1}^N \nabla_i^2 + \sum_{i=1}^N \mathcal{V}_{\rm He-\graphene}(\vec{r}_i) + 
    \sum_{i <j} \mathcal{V}_{\rm He - He}(\vec{r_i}-\vec{r_j}) 
\label{eq:Ham}
\end{equation}
where the $i^{\text{th}}$ atom is located at position $\vec{r}_i=(x_i,y_i,z_i)$ and we have neglected 3-body interactions. The interaction between helium atoms $\mathcal{V}_{\rm He-He}$, shown in Fig.~\ref{fig:VHeGraphene}(a), has been parameterized to reproduce experimental results to high accuracy \cite{Przybytek:2010ol,Cencek:2012iz}, while the corrugated
helium--graphene potential $\mathcal{V}_{\rm He-\graphene}$ can be constructed empirically \cite{Steele:1973fo,Carlos:1979jq, Carlos:1980cf, Vidali:1980pg,Pirani:2001rs, Pirani:2004sc,Bruch:2007bk,Bruch:2010fq,Badman:2018cq}. 
Here, we employ the form of Ref.~\cite{Steele:1973fo}, obtained from the sum of isotropic interactions between $^4$He and C atoms with the 6--12 Lennard--Jones potential with parameters $\sigma$ and $\varepsilon$: 
\begin{multline}
    \mathcal{V}_{\rm He-\graphene}(\vec{r}_i) =  \varepsilon\sigma^2\frac{4\pi}{A} \left\{
\qty[\frac{2}{5}\qty(\frac{\sigma}{z_i})^{10}-\qty(\frac{\sigma}{z_i})^{4}] \right. \\ 
    + \sum_{\vb*{g}\ne 0}\sum_{\ell=1}^{2} \mathrm{e}^{\imath \vb*{g}\cdot(\vec{\mathcalligra{r}}_i-\vec{b}_\ell)} \left[\frac{1}{60}\qty(\frac{g\sigma^2}{2z_i})^{5}K_5(gz_i) \right.  \\
\left.\left. - \qty(\frac{g\sigma^2}{2z_i})^{2}K_2(gz_i)\right]  \right\}.
\label{eq:HeGraphene}
\end{multline}
In Eq.~(\ref{eq:HeGraphene}), $\vrg_i = (x_i,y_i)$ are the coordinates of a $^4$He atom in the $xy$-plane, $\vec{b}_\ell$ are the basis vectors defined in Eq.~(\ref{eq:grapheneBasis}), $\vec{g} = n_1 \vec{G}_1 + n_2 \vec{G}_2$ are the reciprocal lattice vectors with magnitude $g \equiv \abs{\vec{g}}$ where $n_1,n_2 \in \mathds{Z}$,
\begin{equation}
    \vec{G}_1 = \frac{2\pi}{3a_0}\qty(\sqrt{3},1), \qquad 
    \vec{G}_2 = \frac{2\pi}{3a_0}\qty(-\sqrt{3},1),
\label{eq:grapheneG}
\end{equation}
and $A = 3\sqrt{3}a_0^2/2$  is the area of the unit cell. $K_n$ are modified Bessel functions which decay as $\exp(-gz_i)$ at large argument. The parameters $\varepsilon$ and $\sigma$ have been previously calculated for graphene by matching the dispersion force originating from a continuum approximation for its polarizability at large separations to that predicted by Eq.~\eqref{eq:HeGraphene} \cite{Nichols:2016hd}. We use: $\varepsilon = \SI{16.961}{\kelvin}$ and $\sigma=\SI{2.643}{\angstrom}$, which are different from previous studies that employed parameters determined for graphite \cite{Corboz2008cb, Gordillo:2009jb, Gordillo:2014cp,Gordillo:2012fl, Happacher:2013ht}. The resulting empirical potential is shown in Fig.~\ref{fig:VHeGraphene}(b) which has a minimum at the center of a graphene hexagon a distance $z_{\min}\simeq\SI{2.5}{\angstrom}$ above the membrane with depth $\sim\SI{-190}{\kelvin}$. Since the smallest reciprocal lattice vector, $|\vb*{g}_1| \simeq \SI{3}{\angstrom^{-1}}$, $|gz| \approx 7 \gg 1$ near the minimum, and in practice, the sum over $\vec{g}$ converges rapidly such that only a few sets with equal $\abs{\vec{g}}$ need to be retained.
%
%

\subsection{Ab Initio}%
\label{subsec:semiEmpirical}

Here we briefly discuss the main conceptual differences between the ``empirical" approach outlined above, where the van der Waals interactions are used to calculate physical quantities via many-body techniques (such as the Hartree--Fock method and quantum Monte Carlo), and ab initio methods. In the latter, calculations are performed using the usual Born--Oppenheimer approximation where all atomic nuclei are considered classical and only the electrons receive a full quantum-mechanical treatment.  Using van der Waals corrected Density Functional Theory and 2nd order M\o{}ller--Plesset perturbation theory, we have computed the effective interactions between He--He and He--graphene with the results shown in Fig.~\ref{fig:VHeGraphene} with the details included in \S~\ref{subsec:DFT}-\ref{subsec:MP2}.  

For interactions between two helium atoms, MP2 agrees very well with the empirical potential \cite{Przybytek:2010ol,Cencek:2012iz} (as seen in Fig.~\ref{fig:VHeGraphene}(a)) motivating the choice of basis set employed.  The dispersion corrected DFT predicts He--He hard-core interactions that are weaker than $\Vhehe$ at short distances but correctly captures the location of the minimum $r_{\rm min} \simeq \SI{2.75}{\angstrom}$.  The ab initio computation of the height ($z$) dependence of the adsorption potential at a fixed position in the $xy$ plane corresponding to the center of a graphene hexagon, (as seen in Fig.~\ref{fig:VHeGraphene}(b)) yields a value of $z_{\rm min} \simeq 2.5-3~\si{\angstrom}$ where the minima is observed with a depth varying between -400 and \SI{-300}{\kelvin}.

The agreement between the He--He and He--graphene potentials is remarkable, in light of the drastically different approximations at play (e.g. frozen nuclei vs. dispersion) and the large variance in $\Vext$ reported in the literature for various ab initio approaches \cite{Burganova:2016hj}. 

All adsorption potentials lead to the existence of a well-defined monolayer of $^4$He on graphene.  Even with the differences between the interactions on display in Fig.~\ref{fig:VHeGraphene}, the resulting effective 2D low energy model that describes the system will turn out to be remarkably similar.

\section{Dimensionality of the First Adsorbed Layer}%
\label{sec:monolayer}

Regardless of the form of the employed interaction potential in the microscopic model, the goal of this work is to obtain access to properties of the ground state of the $N$-particle three-dimensional time-independent Schr{\"o}dinger equation:
\begin{equation}
    H \Psi_0(\vec{R}) = E_0 \Psi_0(\vec{R}) 
\label{eq:SWE}
\end{equation}
in order to determine the parameters of an effective two-dimensional Bose--Hubbard Hamiltonian described by Eq.~\eqref{eq:BHHamiltonian} where $\vec{R} \equiv\qty{\vec{r}_1,\dots,\vec{r}_N}$ are the spatial locations of helium atoms. 

\begin{figure}[h]
\begin{center}
    \includegraphics[width=\columnwidth]{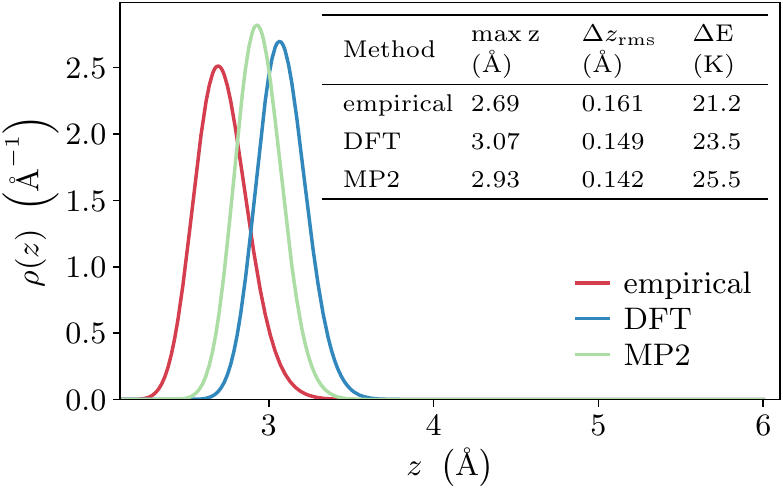}
\end{center}
\caption{The particle density, $\rho(z) \propto |\phi_0(z)|^2$ in Eq.~\eqref{eEmpadd1_01} obtained via the shooting method for $\Vext$ computed for the three different approaches to the potentials described in \S~\ref{subsec:empirical}-\ref{subsec:semiEmpirical}. The table indicates the value of $z$ at which the maximum density occurs, the root mean squared value of $z$ $\Delta z_{\rm rms} = 
\sqrt{\expval{z^2} - \expval{z}^2}$, and the difference between the zero point energy and the minimum of $\Vext$ for each potential. All three methods yield density profiles for the adsorbed layer that are effectively two dimensional with sub-angstrom widths.}
\label{fig:ShootingMethod}
\end{figure}

The basic physical picture of adsorption of helium on graphene is clear.  At low temperature and densities, atoms preferentially adsorb to the strong binding sites located at the center of graphene hexagons due to the attractive interaction seen in Fig.~\ref{fig:VHeGraphene}(b). If the density is low enough that interactions between helium atoms are not relevant, Eq.~\eqref{eq:Ham} can be numerically integrated to obtain the $z$-dependence of the wavefunction in the approximation where the corrugation is neglected and the adatoms experience an average smooth potential over the $xy$-plane (\emph{i.e.}~taking only the $\vb*{g}=0$ term in Eq.~\eqref{eq:HeGraphene}, see Appendix \ref{app:3Dto2D} for details). The resulting single particle density in the $z$-direction \cite{Campbell:1972ct, Whitlock:1998gb, Gordillo:2009jb} is shown in Fig.~\ref{fig:ShootingMethod} along with values corresponding to the adsorption potentials computed via ab initio methods.
Thus,  single atoms are strongly localized around $z \approx \SI{3}{\angstrom}$, regardless of the way the adsorption potential is calculated, with an root mean squared width of $\approx\SI{0.15}{\angstrom}$ and a zero point energy that lifts the ground state $\approx\SI{20}{\kelvin}$ above the classical potential minimum.

As the density of adatoms is increased, there is now a competition between the energy gained due to attraction of the graphene sheet, and the interaction potential between helium atoms, $\mathcal{V}_{\rm He-He}$,  which has an attractive minimum at $r_{\rm min} \approx \SI{3}{\angstrom}$ and eventually becomes repulsive at smaller distances (see Fig.~\ref{fig:VHeGraphene}). The length scales defining $\Vhehe$ should be compared with those imposed by the graphene corrugation potential where the nearest neighbor distance between two hexagon centers is $\rg_{\rm\,NN} = \sqrt{3}a_0 \simeq \SI{2.46}{\angstrom}$ while the next-nearest neighbor distance, corresponding to one out of every three hexagons occupied has $\rg_{\rm\,NNN} = 3a_0 \simeq \SI{4.26}{\angstrom}$ as seen in Fig.~\ref{fig:triangular}. Thus at low densities, the system stabilizes at a single well-defined 2D monolayer, that can exist at both commensurate and incommensurate filling fractions $f=N/N_\graphene$ (where $N_\graphene$ is the number of triangular lattice sites) in a regime of coverage where both the adsorption and interaction energies are attractive.  As the density continues to increase, eventually the cost of repulsive interactions between helium atoms overcomes the reduced attraction felt further from the sheet and layer completion is reached near $f \approx 0.6$.  At this point, a second layer begins to form and the system can no longer be considered as effectively two dimensional (see Fig.~\ref{fig:Eos} in \S~\ref{subsubsec:manybody}).

This simple picture has been validated by 50 years of experiments \cite{Bretz:1971jo,Bretz:1973ky, Dash:1979bk, Zimmerli:1992hz,Dash:1994cp,Zimmerli:1988ii, Shibayama:2009jn, Nakamura:2014vx,Nyeki:2017ef} and numerical simulations \cite{ Whitlock:1998gb, Pierce:1999cn, Pierce:2000cj, Corboz2008cb, Bruch:2010fq, Gordillo:2011jb, Ahn:2016dm, Badman:2018cq,Gordillo:2020wz} on helium adsorbed on graphite, where the adsorption potential is 10\% stronger than graphene.  While no experiments yet exist in the graphene system considered here, quantum Monte Carlo simulations \cite{Gordillo:2009jb,Gordillo:2011jb,Gordillo:2012fl,Kwon:2012ie,Happacher:2013ht,Gordillo:2014cp,Markic:2016dm} both at zero and finite temperature show analogous behavior. As already discussed in the introduction, in the first layer, a commensurate $\sqrt{3} \times \sqrt{3}$~$\mathrm{R}~30^{\circ}$ incompressible C1/3 solid phase (helium atoms occupy 1/3 of the strong binding sites on a triangular lattice (hexagon centers) with constant $\sqrt{3}a_0$ and axes rotated (R) by \SI{30}{\degree} with respect to the original graphene triangular lattice) 
is thermodynamically stable over a large range of chemical potentials \cite{Happacher:2013ht} (see Figs.~\ref{fig:cell} and \ref{fig:triangular}) and may compete with a lower density liquid \cite{Gordillo:2014cp} depending on simulation details and the employed form of $\mathcal{V}_{He-\graphene}$. All observed phases in the first layer are incompressible, with no systematic evidence of finite superfluid density surviving extrapolation to the thermodynamic limit.

Thus, the ground state of the first adsorbed monolayer of $^4$He on graphene can be described by an effective two-dimensional system. We now discuss how it can be mapped at low energies onto a single-band Bose--Hubbard model, which requires moving beyond the simple continuum one-body model described here and understanding the role of interactions in Eq.~\eqref{eq:Ham}. 

\section{Effective 2D Bose--Hubbard Description}
\label{sec:BHParameters}

We attack this problem at various levels of sophistication starting from the non-interacting band structure and Wannier theory (where we analyze the corrugation of the adsorption potential) and systematically explore the effects of interactions in different approximation schemes: Hartree, Hartree--Fock, quantum Monte Carlo, M\o{}ller--Plesset and dispersion corrected density functional theory.  

In this section we introduce an effective 2D $\mathcal{V}_{\rm He-\graphene}(\vrg)$ potential where $\vrg = (x,y)$ is the in-plane coordinate. The way $\mathcal{V}_{\rm He-\graphene}(\vrg)$ is determined depends on the specific method used and will be discussed on a case by case basis.

\subsection{Mapping onto a Bose--Hubbard Model}
\label{subsec:general}

First, we briefly outline the well-known general procedure for mapping the interacting problem in Eq.~\eqref{eq:Ham}, onto the effective  Bose--Hubbard model  Eq.~\eqref{eq:BHHamiltonian}.  This mapping is valid at low energies and therefore the two representations lead to the same ground state properties. A similar mapping has been used to analyze the properties of dilute Bose gases confined on optical lattices \cite{Bloch:2008,Jaksch:1998,Jaksch:2013}; however the physics in our case turns out to be fundamentally different due to the importance of short-range correlations for a (fairly dense) collection of  helium atoms confined to the graphene lattice.

We begin by expressing the first-quantized microscopic Hamiltonian in Eq.~\eqref{eq:Ham} in second quantization for a single 2D monolayer,  via the introduction of bosonic field operators, $\hat{\Psi}(\vrg),\hat{\Psi}^{\dagger}(\vrg)$ such that the local density is $n(\vrg) = \hat{\Psi}^{\dagger}(\vrg)\hat{\Psi}(\vrg)$. In this notation, the effective 2D Hamiltonian can be written as a sum of a one-particle term, which includes the kinetic energy and the helium--graphene potential, and a two-body term (that originates from the helium--helium interaction) \cite{LL9,Mahan}:
\begin{eqnarray}
&H& = \int \dd{\vrg} \hat{\Psi}^{\dagger}(\vrg) \qty( -\frac{\hbar^2}{2m}\nabla_{\vrg}^2 + \mathcal{V}_{\rm He-\graphene}(\vrg))  \hat{\Psi}(\vrg)    + \nonumber \\ 
&& \frac{1}{2} \iint \dd{\vrg}\dd{\vrg^\prime} \hat{\Psi}^{\dagger}(\vrg) \hat{\Psi}^{\dagger}(\vrg') \mathcal{V}_{\rm He - He}(\vrg-\vrg')  \hat{\Psi}(\vrg') \hat{\Psi}(\vrg)   \, , 
\label{eq:Hamsecond}
\end{eqnarray}
where a discussion of $\Vhehe$ is included in Appendix~\ref{app:3Dto2D}.
For helium atoms strongly confined near 2D triangular lattice locations $\vrg_i$ defined by the centers of graphene hexagons (see Fig.~\ref{fig:triangular}), the field operators can be expanded over a complete orthonormal set of localized Wannier functions $\psi(\vrg -\vrg_i )$ and the bosonic annihilation and creation operators $b_{i},b^{\dagger}_i$ \cite{LL9,Mahan}:
%
\begin{equation}
\hat{\Psi}(\vrg) = \sum_{\vrg_i} \psi(\vrg -\vrg_i ) b_i,  \ \
 \hat{\Psi}^{\dagger}(\vrg) = \sum_{\vrg_i} \psi^*(\vrg -\vrg_i ) b^{\dagger}_i .
 \label{eq:fieldops}
\end{equation}
We use the shorthand notation $b^{\dagger}_{i}\equiv b^{\dagger}_{\vrg_i}$ for an operator that creates a boson at $\vrg_i$, and $\psi_{i}(\vrg) = \psi(\vrg -\vrg_i )$ for the Wannier function localized around the site $i$ on a triangular lattice.
The Wannier functions will be constructed in the next section, and we assume that they correspond to the lowest energy band. 
   
Substituting Eq.~\eqref{eq:fieldops} into \Eqref{eq:Hamsecond} one obtains the effective lattice Bose--Hubbard Hamiltonian in Eq.~\eqref{eq:BHHamiltonian}, where the boson density is $n_i=b^{\dagger}_ib_i$.  The one-particle hopping $(t)$ and the density--density interactions on-site $(U)$, at
nearest-neighbor sites $(V)$, and next-nearest-neighbor sites $(V')$ on a triangular lattice are then given by the expressions \cite{Mahan}:
\begin{align}
    \label{eq:tBH}
    t &= -\int \dd{\vrg} \psi^*(\vrg) \qty[-\frac{\hbar^2}{2m}\nabla_{\vrg}^2 + \mathcal{V}_{\rm He-\graphene}(\vrg)]
    \psi(\vrg - \vb*{a}_1) \\
\label{eq:UBH}
U &= \iint \dd{\vrg} \dd{\vrg^\prime}  \abs{\psi(\vrg)}^2 \mathcal{V}_{\rm He - He}(\vrg-\vrg') \abs{\psi(\vrg')}^2
 \\
\label{eq:VBH}
V &= \iint \dd{\vrg}\dd{\vrg^\prime} \abs{\psi(\vrg)}^2 \mathcal{V}_{\rm He -
He}(\vrg-\vrg') \abs{\psi(\vrg' - \vb*{a}_1)}^2  \\
\label{eq:VpBH}
V^\prime &= \iint \dd{\vrg} \dd{\vrg^\prime} \abs{\psi(\vrg)}^2
\mathcal{V}_{\rm He - He}(\vrg-\vrg') \abs{\psi(\vrg' - \vb*{a}_1-\vb*{a}_2)}^2 .
\end{align}
Here, the choice of lattice site for the computation is arbitrary due to translational invariance, and one can replace $\vb*{a}_1 \leftrightarrow \vb*{a}_2$.


\subsection{Band Structure and Effective Hopping $t$}
\label{subsec:bands}

In order to calculate the overlap integrals in Eqs.~\eqref{eq:tBH}--\eqref{eq:VpBH}, we start by evaluating the general band structure and specifically  the hopping parameter $t$ which are  determined by the purely one-particle Hamiltonian, 
$-\frac{\hbar^2}{2m}\nabla_{\vrg}^2 + \mathcal{V}_{\rm He-\graphene}(\vrg)$.
For a given effective 2D  potential $\mathcal{V}_{\rm He-\graphene}(\vrg)$, the procedure is described in the literature \cite{Kaxiras:2019}. 
Bloch's theorem states that the solutions to the Schr\"odinger equation in a periodic potential are the product of a periodic function $u_{\vec{k}}^{(n)} (\vrg)$ and a plane wave, $\Psi_{\vec{k}}^{(n)}(\vrg) = \mathrm{e}^{i \vec{k} \cdot\, \vrg} u_{\vec{k}}^{(n)} (\vrg)$, where $\vec{k}$ is the 2D lattice quasi-momentum.  The index $n$ labels the different bands with corresponding energies $\varepsilon^{(n)}(\vec{k})$.
We seek wave functions for the lowest energy band $(n=1)$ and hence
omit the band index for simplicity. Once the Bloch wave-functions are found, the localized Wannier functions are constructed via
\begin{equation}
\psi(\vrg -\vrg_i ) = \frac{1}{\sqrt{N_\graphene}} \sum_{\vec{k} \in BZ} \mathrm{e}^{-i \vec{k} \cdot\, \vrg_i } \Psi_{\vec{k}}(\vrg) ,
\label{eq:Wannier}
\end{equation}
where the summation is over the first Brillouin zone, and $N_\graphene$ is the number of (triangular) lattice sites. Eq.~\eqref{eq:Wannier} can now be used in the overlap integral for $t$ defined in Eq.~\eqref{eq:tBH} for a given value of $\Vext$ computed within the empirical or ab initio approach.

\subsubsection{Empirical}
Here the bare potential is given by Eq.~\eqref{eq:HeGraphene} and we use two approaches to construct an effective 2D  potential $\mathcal{V}_{\rm He-\graphene}(\vrg)$.

\begin{figure}[t]
\begin{center}
    \includegraphics[width=\columnwidth]{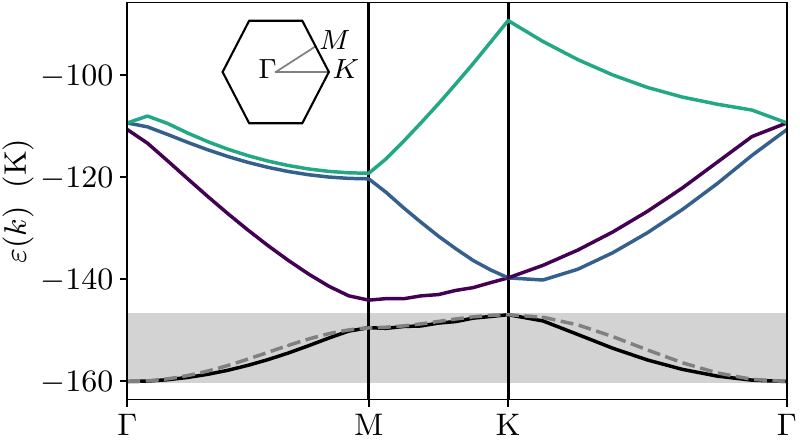}
\end{center}
\caption{The band structure obtained using $\tilVHeG\qty(\vrg)$ as introduced Eq.~\eqref{eEmp1_04c} along a high symmetry path in the first Brillouin zone as shown in the inset. The first band (with $n=1$) is well separated from the higher excited bands and thus the low energy properties of the system are determined by the lowest band. The dashed line shows the tight binding dispersion from Eq.~\eqref{eq:tbdispersion}, in excellent agreement with the continuum model supporting the use of an effective 2D lattice model. The shaded region corresponds to the bandwidth equal to $9t$ by symmetry.}
\label{fig:BandStructure}
\end{figure}

\begin{figure}[h]
\begin{center}
\includegraphics[width=\columnwidth]{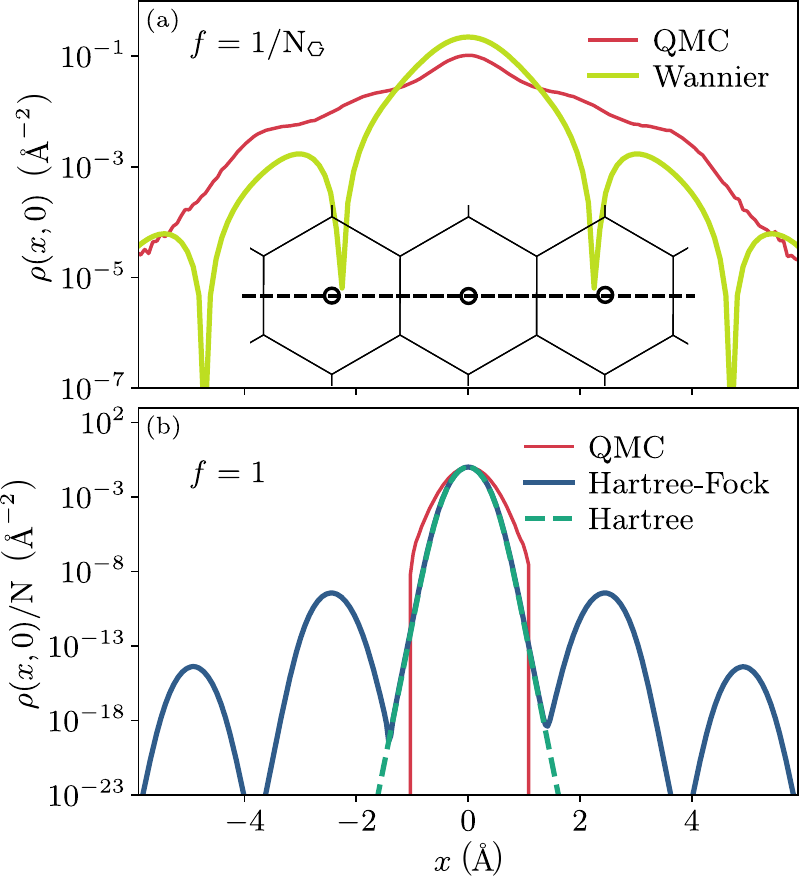}
\end{center}
\caption{Spatial dependence of the density $\rho(x,y) = |\psi(x,y)|^2/N$ of an adsorbed $^4$He atom on graphene for a $y=0$ cut in the $xy$-plane corresponding to the lattice path shown in the top inset. (a) A comparison of the localized Wannier function defined in Eq.~\eqref{eq:Wannier} to that computed via quantum Monte Carlo (QMC) for a single particle $N=1$, $f=1/N_\graphene$ by slightly biasing a single site at the level of the trial wavefunction as discussed in  \S~\ref{subsec:qmc}. The wavefunction strongly penetrates into neighboring lattice sites, leading to the breakdown of Wannier theory for the computation of interaction parameters.  (b) The density at unit filling $N = N_\graphene$ as computed via QMC, and within the Hartree-Fock and Hartree approximations showing the tendency towards exponential localization on a single site. In both panels, the 2D normalization is computed over the full graphene sheet.
}
\label{fig:WannierCut}
\end{figure}

Following the discussion in \S~\ref{sec:monolayer}, we can integrate the full 3D helium--graphene interaction potential over the probability density in the $z$-direction presented in Fig.~\ref{fig:ShootingMethod} as described in detail in Appendix \ref{app:3Dto2D}.  This leads to a 2D potential $\tilVHeG\qty(\vrg)$ as defined in Eq.~\eqref{eEmp1_04c}. The corresponding band structure is presented in Fig.~\ref{fig:BandStructure} and the resulting Wannier function is plotted in Fig.~\ref{fig:WannierCut}(a). 
Based on these results, Eq.~\eqref{eq:tBH} is evaluated in the lowest band resulting in:
\begin{equation}
    t_{\mathrm W} = \SI{1.45}{\kelvin}\, .
    \label{eq:TWannier}
\end{equation}
In addition,  within the tight-binding approximation the lowest band is described by the explicit formula:
\begin{equation}
\varepsilon(\vec{k}) - \varepsilon_0 = \! -2t \!  \qty[\!
\cos(k_xa) + 2\cos(\frac{k_xa}{2}) \cos(\frac{\sqrt{3}k_ya}{2})],
\label{eq:tbdispersion}
\end{equation}
where $a=\sqrt{3}a_0$ and $\varepsilon_0$ is an energy offset.
This means that the bandwidth, defined as the energy difference between the K point 
(located at momentum $(4\pi/3a,0)$) and the $\Gamma$ point  $(0,0)$ 
in Fig.~\ref{fig:BandStructure}, is equal to $9t$. Equation \eqref{eq:tbdispersion} is plotted as the dashed line in Fig.~\ref{fig:BandStructure}, and the considerable agreement provides further validation for mapping from the continuum to a lattice model.

An alternative approach to obtaining a 2D effective potential is to exactly simulate a single $^4$He atom subject to the full 3D potential via quantum Monte Carlo as described in detail in \S~\ref{subsec:qmc} and obtain the adsorption potential as a function of the 2D coordinate in the plane, $\vrg$, via Eq.~\eqref{eq:VHeGrapheneAverage}: $\mathcal{V}_{\rm He-\graphene}(\vrg)=\expval{\mathcal{V}_{\rm He-\graphene}(x,y)}$.
The corresponding hopping parameter
calculated from this potential is
\begin{equation}
 t_{\rm QMC} = \SI{1.38(1)} {\kelvin}. 
 \label{eq:tQMC}
\end{equation}
where the parenthesis indicates the statistical uncertainty in the last digit.  We note that this value agrees with that computed using the adsorption potential determined from the 1D wavefunction in Eq.~\eqref{eq:TWannier} at the order of 10\%.

\subsubsection{Ab Initio}

\begin{figure}[t]
\includegraphics[width=1.00\columnwidth]{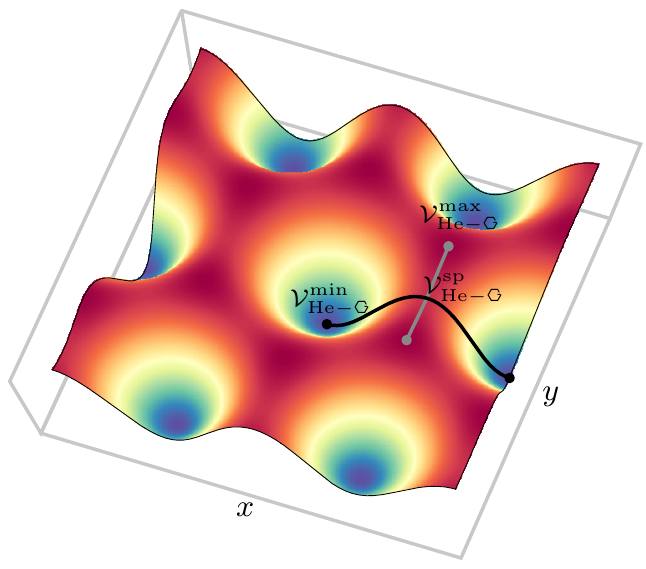}
\caption{The effective 2D potential $\Vext(x,y)$ used to calculate the hopping $t$ can be reconstructed from MP2 and DFT calculations by determining three values corresponding to the minimum, maximum, and saddle-point values as indicated. The resulting scale along the black line can be seen in Fig.~\ref{fig:Tunneling}.}
\label{fig:Steele-Path}
\end{figure}

The hopping parameter $t$ can also be estimated for an effective 2D potential computed within the ab initio approximation.  While it is computationally difficult to perform a DFT and MP2 calculation for every position $\vrg$, these numerical methods can readily determine the adsorption potential at the high symmetry points corresponding to the minima, maxima, and saddle point (as shown in Fig.~\ref{fig:Steele-Path}).  
%
%
Since the summation over $\abs{\vec{g}}$ is dominated by the terms with the smallest magnitudes and converges rapidly, the full 2D potential can be approximated as
\begin{equation}
\mathcal{V}_{\rm{He-\graphene}}(\vrg) = \mathcal{V}_0 + c_{g_1} \sum_{|\vec{g}| = g_1} \mathrm{e}^{i \vec{g}\cdot\, \vrg} + c_{g_2} \sum_{|\vec{g}| = g_2} \mathrm{e}^{i \vec{g} \cdot\, \vrg},
\label{eq:Vsym}
\end{equation}
where  $g_1 = 4\pi/(3a_0)$ and $g_2 =\sqrt{3}g_1$ are the lengths of the two smallest set of $\vec{g}$ vectors . The coefficients 
 $c_{g_{1,2}}$
can be uniquely determined from the minimum, maximum, and saddle point values of the potential as 
\begin{align}
    c_{g_1} &= -\frac{1}{9} (\mathcal{V}_{\rm He-\graphene}^{\rm max} - \mathcal{V}_{\rm He-\graphene}^{\rm min}) ,\nonumber \\
    c_{g_2} &= \frac{1}{8} \qty(\mathcal{V}_{\rm He-\graphene}^{\rm sp} - \mathcal{V}_{\rm He-\graphene}^{\rm min}) - \frac{1}{9}\qty(\mathcal{V}_{\rm He-\graphene}^{\rm max}- \mathcal{V}_{\rm He-\graphene}^{\rm min})\, .
\end{align}
A summary of the relevant parameters calculated with different methods is presented in Table \ref{tab:VHeGraphene}.
\begin{table}[h]
    \renewcommand{\arraystretch}{1.5}
    \setlength\tabcolsep{12pt}
    \begin{tabular}{@{}lll@{}} 
        \toprule
        Method & $\Vext^{\rm max} - \Vext^{\rm min} \; (\si{\kelvin})$ & 
        $\Vext^{\rm sp} - \Vext^{\rm min} \; (\si{\kelvin})$ \\
        \midrule
        HF & 21.2 & 17.5 \\
        QMC & 24.7 & 21.7 \\
        DFT  & 39.2 & 36.1 \\
        MP2  & 72.2 & 66.0 \\
        \bottomrule
    \end{tabular}
    \caption{\label{tab:VHeGraphene} The parameters taken from the adsorption potential for the four different methods at the high symmetry points corresponding to the minima, maxima, and saddle point required to calculate the coefficients $c_{g_{1,2}}$ in \Eqref{eq:Vsym} (HF and Wanner use the same potential). }
\end{table}

The calculation of the hopping parameter $t$ then proceeds as in the previous section, where the Wannier functions are determined using the 2D potential in \Eqref{eq:Vsym}, leading to:
\begin{equation}
   t_{\rm DFT} = \SI{1.10}{\kelvin}, \   t_{\rm MP2} = \SI{0.59}{\kelvin} \, .
\end{equation}

The results for $t$ from the QMC, and DFT adsorption potentials are remarkably similar given the variation in their underlying approximations to the full 3D system. MP2, on the other hand, predicts a smaller value of $t$, as a result of the significantly stronger adsorption potential $\Vext(\vrg)$ from this method. 

As a final check on the physical realism of these results, the WKB method can be used to estimate $t$ as discussed in Appendix~\ref{app:WKB}, leading
to results which are in very reasonable agreement with those presented above.

\subsection{Interaction Effects: Breakdown of the Wannier Theory}
\label{subsec:breakdown}

So far, we have been considering the mapping of the 2D adsorbed $^4$He layer within the single particle approximation. Now, we proceed with an evaluation of the interaction parameters $U,V,V'$. In any parameterization of the He--He interaction potential, the existence of a strong hard-core will preclude the double occupation of a single site on the triangular lattice and thus effectively $U = \infty$ as it is the dominant scale with $U \gg t, V,V^\prime$. For the potentials in Fig.~\ref{fig:VBHqmc}(a) we find that Eq.~\eqref{eq:UBH} yields $U > \SI{E6}{\kelvin}$.  Therefore, the effective Bose--Hubbard model describes hard-core bosons hopping on the triangular lattice formed by the graphene hexagon centers (Fig.~\ref{fig:triangular}).  Using the single-particle Wannier function approach,  one can also  compute the nearest neighbor ($V$) and next-nearest neighbor ($V^\prime$) parameters directly from Eqs.~\eqref{eq:VBH} and \eqref{eq:VpBH} which lead to  
\begin{equation}
    V_{\rm W} = \SI{7540}{\kelvin}, \   V_{\rm W}' = \SI{638}{\kelvin} .
\label{eq:naiveV}
\end{equation}

The resulting enormous energy scales associated with these parameters are unphysical and suggest that the spatial extent of one-particle wave function is too large, and fails to capture the correct interaction physics. This catastrophe originates from the fact that we study the adsorption of $^4$He atoms on a solid-state substrate and consequently
both the spatial extent of the one-particle wavefunction (determined by the graphene lattice structure), and the most prominent (repulsive) part of the He--He potential, vary on the same length scale, of order several \AA.
 As outlined in the Introduction, this behavior is in contrast with  cold atomic gases confined in optical
lattices where the mentioned length scales are well separated, leading to a much simpler,  finite  $U$ Hubbard model \cite{Bloch:2008} with irrelevant interactions $V,V'$.



The very
strong He--He repulsion on the scale of the one-particle wave-function
effectively produces an infinite on-site Hubbard $U$ and therefore it is the
nearest-neighbor $V$ and next-nearest neighbor $V'$ that determine the relevant quantum phases of the system, leading to the hard-core $t-V-V'$ model considered here.
Therefore, the determination of $V,V'$  presents considerable technical challenges and has to be done via sophisticated techniques that take into account the correct structure of the wave function which is modified by two-body interactions and at finite density deviates significantly from the one-particle results presented so far. In this sense, our analysis is very different form  the conventional approaches to the Bose--Hubbard model. Because of the well-localized structure of the many-body wave functions (as will be clear from the results of the next sections), the effective Bose--Hubbard model
is still dominated by two-body (density--density) interactions, with the nearest-neighbor term being the largest one ($V\gg|V'|$). 

The remainder of this section presents a number of different approaches to gain access to the many-body wavefunctions of $^4$He on graphene in order to compute $V$ and $V^\prime$ exemplifying the strong correlations in the problem.

\subsection{Hartree--Fock Approach to Interaction Parameters}
\label{subsec:HF}

Here we provide details on how the parameters $V$ and $V^\prime$ of the
effective Bose--Hubbard model can be computed from an effective 2D model of the
adsorbed layer.  Since $V$ is the energy of nearest-neighbor interaction, then
for its computation, one needs to consider a helium layer with a unit filling
fraction. However, as noted in \S~\ref{sec:monolayer}, $^4$He atoms at this
density form two, not one, layers on top of graphene. To resolve this issue, we
will rely on an important result from our QMC simulations, which is described
in detail in \S~\ref{subsec:qmc}. Namely, a quasi-2D, single-layer
arrangement of helium over graphene is restored when one imposes a confining
potential in the $z$-direction.  Importantly, the particle density in that
direction obtained with the confinement is close to the density profile at
filling fraction $f=1/3$; see e.g. Figs.~\ref{fig:chi2} and \ref{fig:Eos}.  This justifies the use of a 2D model for the approximate computation of nearest-neighbor He--He interactions.

Let us stress again that the spatial extent of the maximally localized Wannier functions found in the previous subsection (well-suited for the description of an isolated helium atom), is on the order of the spacing between the nearest graphene hexagon centers. Therefore, the standard approach of computing interaction parameters in the Bose--Hubbard model via the overlap integral Eq.~\eqref{eq:VBH} would give an unphysically large result. However, we note that the mutual repulsion of adjacent helium atoms narrows their wavefunctions considerably compared to the Wannier functions (see Fig.~\ref{fig:WannierCut}). We will now show how these 
narrower wavefunctions are found and then use them in the calculation of $V$ and $V^\prime$ via Eqs.~\eqref{eq:VBH} and \eqref{eq:VpBH}. 

Such wavefunctions are obtained by numerically solving a system of 2D Hartree--Fock equations \cite{PethickSmith:2008}:
\begin{multline}
    -\frac{\hbar^2}{2m}\nabla^2_{\vrg} \psi_i(\vrg) + 
{\mathcal V}_{\rm He-\graphene}(\vrg) \psi_i(\vrg) \\
+ \sum_{i\neq j} \int \dd{\vrg'}\,\psi_j^\ast(\vrg') \mathcal{V}_{\rm He - He}(\vrg-\vrg') \\
\times \left[ \psi_j(\vrg')\psi_i(\vrg) + \psi_i(\vrg')\psi_j(\vrg) \right] 
= \tilde{E}_i \psi_i(\vrg)\,,
 \label{eHF_01a}
\end{multline}
where the wavefunctions $\psi_{i}(\vrg) \equiv \psi(\vrg -\vrg_i )$ 
also satisfy the orthonormality constraint:
\be
\int \dd{\vrg} \psi_i(\vrg) \psi_j(\vrg) = \delta_{ij}\,,
\label{eHF_01b}
\ee
with $\delta_{ij}$ being the Kronecker delta. 

Equations \eqref{eHF_01a}, \eqref{eHF_01b}, were solved by the accelerated imaginary-time evolution method (a variant of fixed-point iterations), whose general framework for systems of equations subject to constraints were laid out in \cite{MultiITEM:2011}; its technical details will be described elsewhere.  To estimate the significance of the exchange interaction (i.e., the last term, $\psi_i(\vrg')\psi_j(\vrg)$, in \Eqref{eHF_01a}), we also simulated the Hartree approximation, obtained from \Eqref{eHF_01a} by dropping that term and not imposing the constraint in \Eqref{eHF_01b}.

We performed simulations for the $z$-averaged potentials ${\mathcal V}_{\rm He-\graphene}$ and ${\mathcal V}_{\rm He-He}$, as described in Appendix~\ref{app:3Dto2D}. 
For the potentials averaged with two different $\rho(z)$'s: that defined in Appendix A and that found by QMC (\S{}~\ref{subsec:qmc}), Eq.~\eqref{eq:VBH} gives, respectively: \ $V_{\rm HF}=69.7$ and $62.2$ K.   The reason for the latter value being smaller is that ${\mathcal V}_{\rm He-He}$ is reduced (smoothened) more by the more spread-out $\rho(z)$ obtained by the QMC. On the other hand, the contribution of the difference between the two averaged ${\mathcal V}_{\rm He-\graphene}$'s to the difference in the corresponding $V$'s is negligible. 
In fact, we found that the effect of even larger --- on the order of 50\% --- changes in the magnitude of ${\mathcal V}_{\rm He-\graphene}$ on $V$ was well under 1\%.
For completeness, we also note that when we used ${\mathcal V}_{\rm He-\graphene}$ and ${\mathcal V}_{\rm He-He}$ averaged with $\rho(z)$ defined in Appendix A but used the Hartree rather than Hartree--Fock approximation, we found 
$V=72.4$ K. 
Finally, the parameter $V'$ computed from \Eqref{eq:VpBH} by any of these approximations equals $-2.1$ K to two significant figures. (The number quoted in Table \ref{tab:BHresults} is for the first aforementioned case.) 

We conclude that  the  Hartree--Fock method leads to remarkably strong  downward renormalization with respect 
to the one-particle  (Wannier theory) result Eq.~(\ref{eq:naiveV}). To summarize, the interaction parameters computed by the Hartree--Fock approximation are: 
\begin{equation}
    V_{\rm HF} = \SI{69.7}{\kelvin}, \   V_{\rm HF}' = -\SI{2.08}{\kelvin} .
\label{eq:HFresults}
\end{equation}
For a system of localized bosons with strong short-range interactions, the Hartree--Fock equations provide a very accurate  description  as many-particle correlations beyond the scope of the method are expected to be weak. In addition, and quite reassuringly, we find that the above results are similar to those obtained by the accurate many-body quantum Monte Carlo technique.

\subsection{Quantum Monte Carlo}
\label{subsec:qmc}

At $T=\SI{0}{\kelvin}$, the path integral ground state quantum Monte Carlo
(QMC) algorithm \cite{Sarsa:2000tt,Cuervo:2005uu,Yan:2017fs} provides access to
ground state properties of a many-body system by statistically sampling the
imaginary time propagator $e^{-\beta H}$. Starting from a trial wave function
$\ket{\Psi_{\mathrm{T}}}$, in the long imaginary time limit
$\beta\rightarrow\infty$, $e^{-\beta H} \ket{\Psi_{\mathrm{T}}}$ converges to
the exact ground state, $\ket{\Psi_0}$, provided $\braket{\Psi_0}{\Psi_\mathrm{T}} \ne 0$.  Within this framework we can directly compute ground state properties by statistically sampling the expectation value of an observable $\mathcal{O}$, via:
\begin{equation}
    \expval{\mathcal{O}} \simeq \frac{\expval{\mathrm{e}^{-\beta H}\mathcal{O}\mathrm{e}^{-\beta H}}{\Psi_T}}{\expval{\mathrm{e}^{-2\beta H}}{\Psi_T}}.
\label{eq:expectation}
\end{equation}
We work in a first-quantized representation $\ket{\vec{R}}$ in 3 spatial dimensions where configurations are sampled from the $3+1$ dimensional imaginary time worldlines of interacting particles.  Appendix~\ref{app:QMCscaling} provides additional details on the convergence and scaling of our QMC approach and the source code can be found online \cite{delmaestroCode}. 

In the remainder of this subsection we discuss how QMC simulations of the 3D
microscopic many-body Hamiltonian in Eq.~\eqref{eq:Ham} can be analyzed in the
context of an emergent 2D Bose--Hubbard model.  We begin by confirming the single-particle description of the adsorbed monolayer described in Section~\ref{sec:monolayer} which allows us to compute an effective 2D potential that can be used to determine the hopping parameters $t$.  We then proceed by reducing the size of the simulation cell in the $z$-direction where the extra dimensional confinement allows us to stabilize a monolayer at the large filling fractions needed to determine the interaction parameters $V$ and $V^\prime$.

\subsubsection{Single Particle Properties: $f = 1/N_\graphene$}

\begin{figure}[t]
\begin{center}
    \includegraphics[width=\columnwidth]{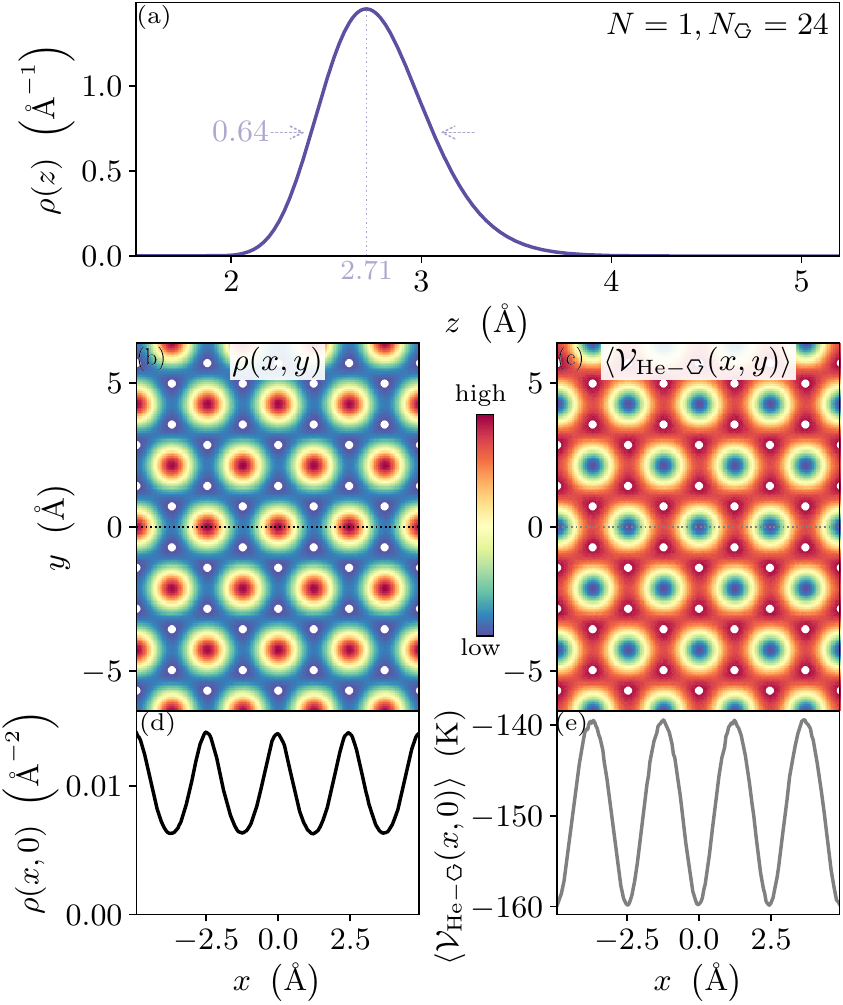}
\end{center}
\caption{Quantum Monte Carlo results for a single $^4$He atom ($N=1$) above a
    graphene membrane with $N_\graphene=24$ adsorption sites.  (a) The average density in the $z$-direction showing a well defined particle position a distance \SI{2.71}{\angstrom} above the sheet with a width of \SI{0.64}{\angstrom}. 
(b) The average density in the $xy$-plane showing the ability of a single particle to \emph{hop} between the sites of the triangular lattice.
White dots show the location of carbon atoms (not to scale). 
(c) The average potential energy experienced by the $^4$He atom due to the graphene sheet in the $xy$-plane.  (d) A horizontal cut of the particle density $\rho(x,y)$ along the line $y=0$. (e) A horizontal cut of the adsorption potential 
$\mathcal{V}_{\rm He-\graphene}(x,y)$ along the line $y=0$.}  
\label{fig:Neq1}
\end{figure}

We begin with the simplest case of considering a single $^4$He atom proximate
to the graphene surface at $T=0$.  The results of QMC simulations are shown in
Figure~\ref{fig:Neq1} for $N=1$ with $N_\graphene = 24$ adsorption sites that
are commensurate in a cell with volume  $L_x \times L_y \times L_z =
\SI{9.84}{\angstrom}\times\SI{12.78}{\angstrom}\times\SI{10.0}{\angstrom} = \SI{1257}{\angstrom^3}$. 
%
%
The cell has periodic boundary conditions in the $x$ and $y$ directions, while motion in the $z$-direction is restricted through the graphene sheet at $z=0$ and a hard wall at $z=L_z$ enforced by the potential
\begin{equation}
    \mathcal{V}_{\rm wall}(z) = \frac{\mathcal{V}_{\rm He -\graphene}(\vb*{r}_{\rm min})}{1 + 
    \mathrm{e}^{(L_z-r_{\rm vdW}-z))/\Delta}}.
\label{eq:Vwall}
\end{equation}
Here, $\vb*{r}_{\rm min} = a_0(\sqrt{3}/2,1/2,1)$ is located at $z=a_0$ above a carbon atom such that $\mathcal{V}_{\rm He -\graphene}(\vb*{r}_{\rm min}) \sim \mathrm{O}(10^5)~\si{\kelvin}$ sets the scale of the repulsive potential, $r_{\rm vdW} \approx \SI{1.4}{\angstrom}$ is the van der Waals radius of helium, while $\Delta = \SI{0.05}{\angstrom}$ defines the rapidness of its onset.  The functional form of Eq.~\eqref{eq:Vwall} and the choice of parameters are unimportant at filling fractions $f \lesssim 1/2$ provided $L_z \gtrsim \SI{6}{\angstrom}$. For the value $L_z = \SI{10}{\angstrom}$ considered here, simulation results are independent of $L_z$ and can be considered to be reflective of bulk adsorption phenomena.

Figure~\ref{fig:Neq1}(a) shows the particle density in the $z$-direction
determined from the expectation value $\rho(z) = \expval{\sum_{i=1}^N
\delta(z_i - z)} \propto \iint \dd{x}\dd{y}\abs{\Psi_0(x,y,z)}^2$ via
Eq.~\eqref{eq:expectation} ($N=1$ here). It has a well-defined peak near
$\SI{2.7}{\angstrom}$ and a corresponding sub-\si{\angstrom} width (shown as
the full width half maximum) demonstrating that adsorbed $^4$He atoms indeed
form a quasi-two dimensional layer.  Panel (b) includes the average particle
density in the plane normalized such that $N = \iint \dd{x}\dd{y}\rho(x,y)$ and
the existence of density in each of the $N_\graphene = 24$ adsorption sites is evidence of particle hopping and an ergodic simulation.  The lower panel (d) is a cut showing the scale of density fluctuations.  Panel (c) shows the average adsorption potential experienced by the $^4$He as it moves in 2D: 
 \begin{equation}
     \expval{\mathcal{V}_{\rm He-\graphene}(x,y)} \equiv \expval{\frac{\int \dd{z} \mathcal{V}_{\rm He-\graphene}(x,y,z)\rho(x,y,z)}{\int \dd{z} \rho(x,y,z)}}
 \label{eq:VHeGrapheneAverage}
 \end{equation}
 while (e) is a horizontal cut along the line $y=0$ highlighting that the minimum-to-saddle corrugation is $\Vext^{\rm sp}-\Vext^{\rm min} \simeq \SI{20.5}{\kelvin}$ (on the order of the kinetic energy).  The trough-to-maximum depth of the adsorption potential is $\Vext^{\rm sp}-\Vext^{\rm min} \simeq\SI{23.6}{\kelvin}$.  These values are reduced by approximately 25\% with respect to the bare potential in Eq.~\eqref{eq:HeGraphene} integrated over the wavefunction in panel (a).  This softening is due to the spatial extent in the $z$-direction and partial localization of the wavefunction in the $xy$-plane.

These QMC results for a single particle can be used in conjunction with the
band structure analysis introduced in \S{}~\ref{subsec:bands} to map the system to a non-interacting Bose--Hubbard model.  In particular, under the assumption that an adsorbed $^4$He atom is confined in a 2D layer, we employed $\expval{\mathcal{V}_{\rm He-\graphene}(x,y)}$ and extracted $t$ from the resulting spectrum in Fig.~\ref{fig:BandStructure}.  This is equivalent in principle to using the overlap in Eq.~\eqref{eq:tBH} for a real wavefunction $|\Psi_\perp(x,y)|^2 \propto \rho(x,y)$ where the QMC average has been performed by exploiting translational invariance, i.e. moving from the Bloch to Wannier basis. The resulting localized single particle wavefunction (labelled QMC) was previously shown in Fig.~\ref{fig:WannierCut}.  We find:
\begin{equation*}
    t_{\rm QMC} = \SI{1.38(1)} {\kelvin}\, .
\end{equation*}

\subsubsection{Many-Body Adsorption: $f>0$}
\label{subsubsec:manybody}

In order to investigate the effects of He--He interactions and thus determine the effective parameters $V$ and $V^\prime$ in the Bose--Hubbard model we need to increase the filling fraction until $^4$He atoms occupy every site of the triangular lattice defined by hexagon centers.  However, as discussed in \S{}~\ref{sec:monolayer}, as the density of helium atoms near the surface is increased, the strong repulsive interaction in Eq.~(\ref{eq:Ham}) will cause layer completion and promote the growth of further layers such that the system can no longer be considered within the 2D approximation.

\begin{figure}[t]
\begin{center}
    \includegraphics[width=\columnwidth]{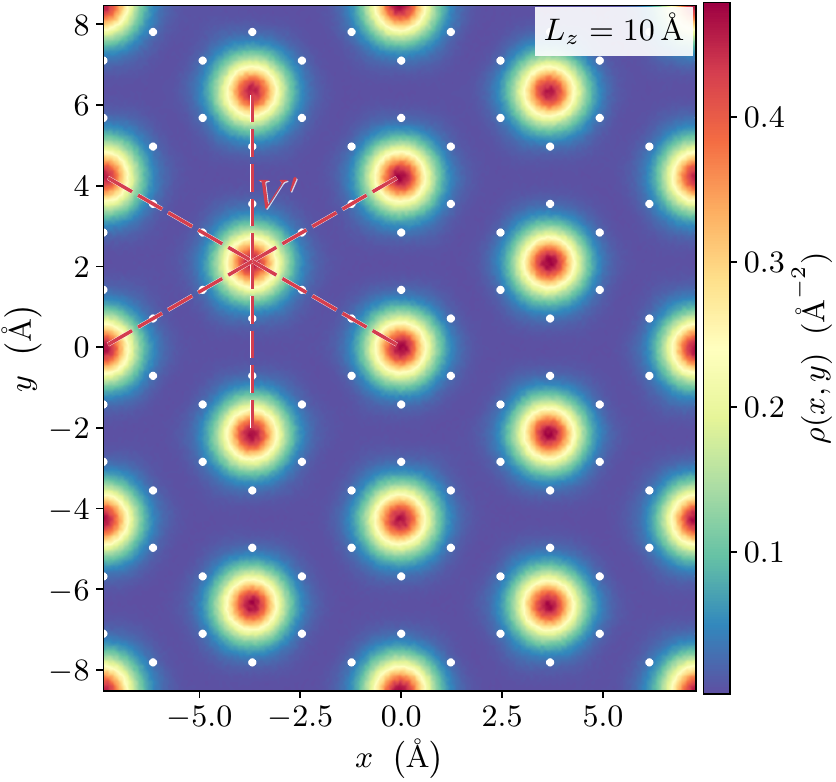}
\end{center}
\caption{The two-dimensional density of particles $\rho(x,y)$ obtained from
ground state quantum Monte Carlo simulations for a simulation cell with
$L_x\times L_y \times L_z = 14.75707 \times 17.04 \times 10.0~\si{\angstrom}$
corresponding to the C1/3 phase with $f=1/3$ for $N=16$ $^4$He atoms on
$N_\graphene = 48$ adsorption sites. Finite size effects in the spatial
wavefunction are negligible beyond $N_\graphene = 12$.}
\label{fig:C13phase}
\end{figure}

We thus begin with the case of $f=1/3$ where a commensurate (C1/3) solid phase is stable over a range of chemical potentials.  Performing a simulation for a system with $N=16$ particles near $N_\graphene = 48$ adsorption sites yields the 2D density profile $\rho(x,y)$ shown in Fig.~\ref{fig:C13phase}.
%
%
Note that in contrast to Fig.~\ref{fig:Neq1}(b) for $f=1/N_\graphene$,  here the local spread of the wavefunction around the hexagon centers in the $xy$-plane is strongly reduced with vanishing density between.  The ground state is a stable solid and interactions are mediated through next-nearest neighbor sites at a distance of $3a_0$ as indicated with dashed lines in analogy with Fig.~\ref{fig:triangular}.  In order to estimate the value of $V^\prime$ from this data, we can compute the ground state energy in the 2D Bose--Hubbard model in Eq.~\eqref{eq:BHHamiltonian} for a Fock state characterizing the C1/3 phase, denoted by $\ket{\cthird}$, where the kinetic energy and nearest neighbor interaction terms are identically zero:
\begin{equation}
    \expval{H_{\rm BH}}{\cthird} \equiv \eval{E_{\rm BH}}_{f=1/3} = 3NV^\prime
    = N_\graphene V^\prime\, ,
\label{eq:E0BHC13}
\end{equation}
and the effects of further $V''$ interactions are neglected.  Measuring the total contribution of the interaction potential to the ground state energy in QMC, $\expval{\Vhehe}$ and equating this with $E_{\rm BH}$, we identify:
\begin{equation}
    V^\prime_{\rm QMC} = \frac{1}{N_\graphene}\expval{\Vhehe}_{f=1/3}
\label{eq:VpQMC}
\end{equation}
and find: 
\begin{equation*}
    V_{\rm QMC}^\prime = \SI{-2.76(2)}{\kelvin}   
\end{equation*}
from the finite size scaling analysis described in Appendix~\ref{app:QMCscaling}.  This value differs by 40\% from the estimate computed from the bare He--He interaction: $V^\prime_{\rm He-He} = \mathcal{V}_{\rm He-He}\qty(\abs{\vec{r}} = 3a_0) \simeq \SI{-2.0}{\kelvin}$.

\begin{figure}[t]
\begin{center}
    \includegraphics[width=\columnwidth]{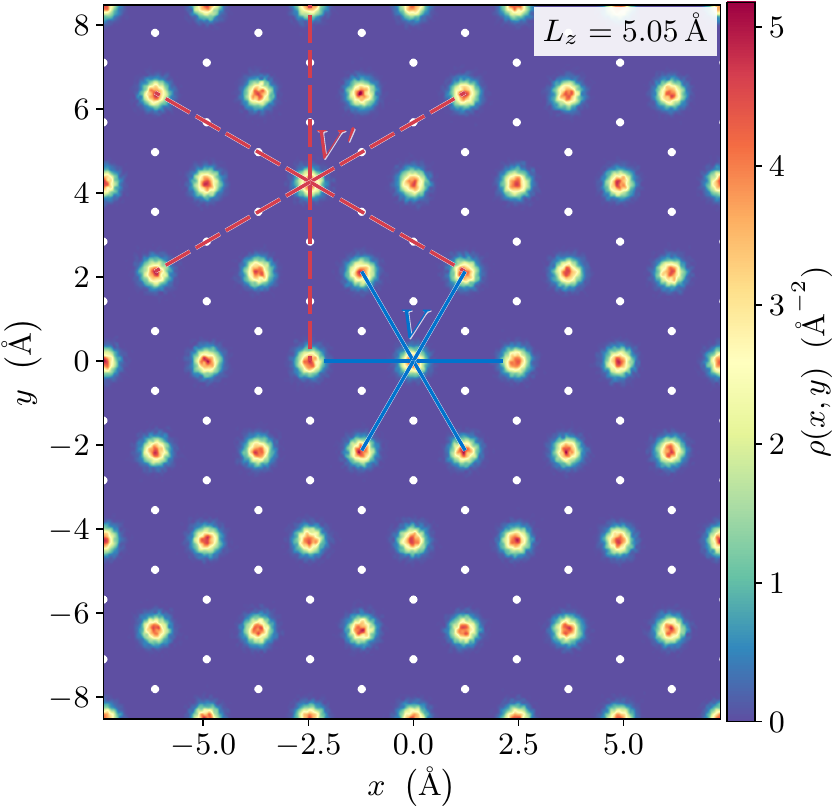}
\end{center}
\caption{The two-dimensional density of particles $\rho(x,y)$ obtained from
ground state quantum Monte Carlo simulations for a simulation cell with
$L_x\times L_y \times L_z = 14.75707 \times 17.04 \times 5.05~\si{\angstrom}$
corresponding to the fully filled phase with $f=1$ for $N=48$ $^4$He atoms on
$N_\graphene = 48$ adsorption sites. Finite size effects in the spatial wavefunction
are negligible beyond $N_\graphene = 12$. Nearest neighbor ($V$) and next-nearest
neighbor ($V^\prime$) couplings in the effective Bose--Hubbard description are
indicated with solid and dashed lines, respectively.}
\label{fig:C1phase}
\end{figure}

\begin{figure}[h]
\begin{center}
    \includegraphics[width=\columnwidth]{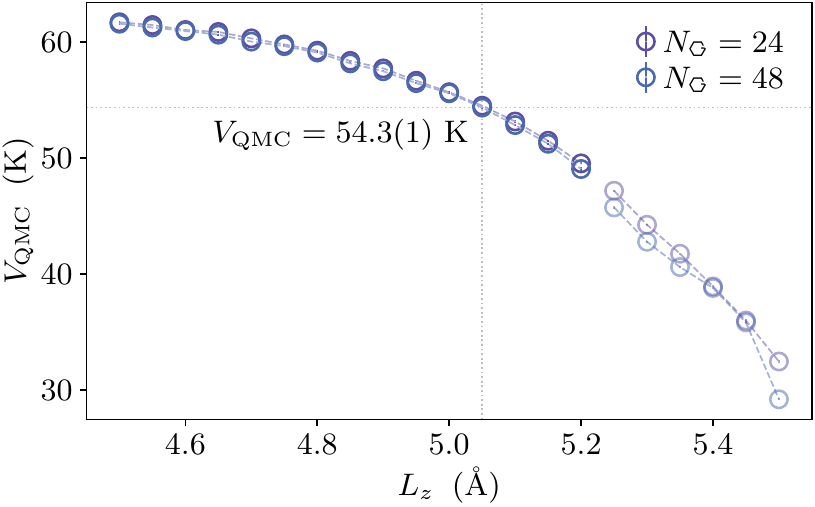}
\end{center}
\caption{The effective nearest neighbor interaction parameter of the Bose--Hubbard model computed from \Eqref{eq:Vqmc} via quantum Monte Carlo for simulation cells with $N_\graphene = 24,48$ as a function of the cell size in the $z$-direction, $L_z$.  The indicated value of  $V_{\rm QMC}=\SI{54.3(1)}{\kelvin}$ was computed at $L_z = \SI{5.05}{\angstrom}$ as described in the text.  The semi-transparent symbols for $L_z \ge \SI{5.25}{\angstrom}$ indicate cell sizes which allowed the nascent formation of a second layer, where the mapping of the microscopic Hamiltonian to the 2D Bose--Hubbard model breaks down.}
\label{fig:VBHqmc}
\end{figure}

In order to perform a similar procedure to extract $V$, we need to hinder the
formation of multiple layers which can be accomplished by restricting our
simulation cell in the $z$-direction using \Eqref{eq:Vwall}.  However,  it is
not clear which value of $L_z$ will (1) maintain the existence of a single
well-defined 2D monolayer as the filling is increased past $f\simeq 0.6$ and
(2) not significantly modify the behavior near filling fraction $f=1/3$ where
the equation of state shows a minimum. The latter is especially important as the
behavior of the 2D Bose--Hubbard model is well understood in this regime
\cite{Wessel:2005ik,Gan:2007zd,Zhang:2011iz}.  In order to answer these questions in an unambiguous manner we have performed an extensive analysis of the capped simulation cell with details provided in Appendix~\ref{subsec:singleLayer}.
We find that $L_z = \SI{5.05}{\angstrom}$ is an appropriate choice for simulations at $f=1$, and in this case, the ground state is an insulator as seen in the 2D density in Fig.~\ref{fig:C1phase}.
%
%
Here, particle wavefunctions are strongly localized near the center of graphene
hexagons, and a cut along $y=0$ was previously shown in
Fig.~\ref{fig:WannierCut}. Following similar logic to that employed for the insulating phase at $f=1/3$, \Eqref{eq:VpQMC}, we examine the Bose--Hubbard model on the triangular lattice at $f=1$ where $\expval{H_{\rm {BH}}}{\cfull} \equiv \eval{E_{\rm BH}}_{f=1} = 3 V N + V^\prime N$ and compute
\begin{equation}
    V_{\rm QMC} = \frac{1}{3N}\expval{\Vhehe}_{f=1}  - V^\prime_{\rm QMC}.
\label{eq:Vqmc}
\end{equation}
The results are shown in Fig.~\ref{fig:VBHqmc} 
%
%
as a function of $L_z$ and we identify: 
\begin{equation*}
    V_{\rm QMC}=\SI{54.3(1)}{\kelvin}   
\end{equation*}
at $L_z = \SI{5.05}{\angstrom}$, where the uncertainty in the last digit arises from a combination of stochastic errors and finite size effects. This value is larger than an estimate obtained from the bare interaction potential for two helium atoms separated by the nearest-neighbor distance:  $V_{\rm He-He} = \mathcal{V}_{\rm He-He}\qty(\abs{\vec{r}} = \sqrt{3}a_0) \simeq \SI{31}{\kelvin}$.  While there are very limited finite size effects in $N_\graphene$, the chosen value of $L_z$ does have an effect on the value of $V_{\rm QMC}$,  reducing it from \SI{61.5}{\kelvin} at $L_z = \SI{4.5}{\angstrom}$ to \SI{49.0}{\kelvin} at $L_z = \SI{5.2}{\angstrom}$.  For larger values of $L_z$, there is no longer a single well-defined monolayer, and the rapid reduction in $V$ observed in Fig.~\ref{fig:VBHqmc} can be attributed to the promotion of a second layer where $^4$He atoms can now move to larger values of $z$ to minimize their repulsive interaction (as seen in Fig.~\ref{fig:layering} in Appendix~\ref{subsec:singleLayer}).

\subsection{Density Functional Theory}
\label{subsec:DFT}
We performed DFT calculations with the PBE (Perdew--Burke--Ernzerhof) generalized gradient approximation \cite{PBE} for the exchange--correlation functional and projector augmented wave (PAW) \cite{PAW} pseudopotentials (PPs), as implemented in the Quantum Espresso electronic structure package \cite{QE09,QE17}. For He--graphene calculations, one or two He atoms are placed at a specified distance from a periodic graphene sheet consisting of $6 \times 6$ unit cells within a hexagonal simulation cell with a vacuum region of $\SI{30}{\angstrom}$. For He--He calculations, two He atoms are placed at a specified distance within a cubic simulation cell of $\SI{30}{\angstrom}$. PAW PPs for C and He were obtained from the standard solid-state PP library \cite{SSSP,PSLibrary,HePP}. We applied the DFT-D4 semi-empirical dispersion correction \cite{D4VdW17,D4VdW19,D4VdW20} when computing single point energies and structural optimizations to account for long-range electronic--correlation effects. The energy cut-off for wavefunctions was $\SI{50}{Ry}$ ($\SI{680}{eV}$) and $\SI{360}{Ry}$ ($\SI{4900}{eV}$) for the charge density and potential. The Brillouin zone is sampled using a Monkhorst--Pack grid with $6\times6\times1$ k-points. 

To obtain the energies of the He-graphene interaction along the path that connects two neighboring minima of the potential (between centers of neighboring lattice sites and passing through the saddle point), the position of the He atom is fixed in the plane of the sheet and the optimal distance from the sheet is then found at each point to compute the energy along the minimum energy surface (see Fig.~\ref{fig:Tunneling}). We followed the same approach to find the maximum value of he potential (centered at the position of a C atom) with the results shown in Table~\ref{tab:VHeGraphene}.

For He--He on graphene calculations, two He atoms are placed at the centers of various lattice sites and at an optimal distance from the sheet, obtained beforehand for a single He atom ($z_{\mathrm{opt}}\simeq \SI{3.036}{\angstrom}$, see Fig.~\ref{fig:VHeGraphene}).  The resulting interaction (relative to non-interacting adsorbed atoms) provides an estimate for the nearest and next-nearest neighbor values:
\begin{equation}
    V_{\rm DFT} = \SI{21.4}{\kelvin}\, \quad V_{\rm DFT}^\prime = \SI{-1.36}{\kelvin}.
\label{eq:VBHDFT}
\end{equation}
%

\subsection{M\o{}ller--Plesset Perturbation Theory}
\label{subsec:MP2}

Because the He--He and He--graphene interactions are dominated by dispersion terms which require accurate treatment of the correlation energy \cite{Cramer,Bartlett}, second-order M\o{}ller--Plesset (MP2) \cite{Moller:1934hc} perturbation theory calculations, which in most cases capture ca. 95\% of the correlation energy \cite{Cramer}, were performed using Gaussian 09 \cite{g09} utilizing Pople-type \cite{Pople} bases sets up to 6-31++G(d,3p), which include diffusion of all orbitals, and polarization functions d and p for all atoms.  

Such a high-order basis set was needed to obtain the He--He interactions in
vacuum to reasonable accuracy (Fig. \ref{fig:VHeGraphene}(a)).  To model the
interaction of He atom(s) with graphene (and possible modifications of the
He--He potential on graphene), a sequence of increasing aromatic molecules was
considered (benzene, coronene, hexabenzocoronene, circumcoronene --- the latter
with 54 C and 18 H atoms). The energy of the system was computed for different
values of $z$ between the He atom(s) and the C plane, and the asymptotic energy
for $z \rightarrow \infty$ was removed as a baseline.  To reproduce graphene,
the aromatic molecules were constructed with C--C distances constrained to $a_0 =
\SI{1.42}{\angstrom}$, and only the coordinates of terminating H atoms were optimized. Figure~\ref{fig:VHeGraphene}(b) shows the potential energy vs. height for a single He atom above the center of a circumcoronene molecule. We observed that the calculations converge after hexabenzocoronene and there was a relatively small ``radial dependence'' of $\mathcal{V}_{\rm{He-\graphene}}(\vrg,z)$ for other hexagon centers, making this a reasonable model for He on graphene. 

We also performed scans of the potentials over different positions over the circumcoronene.  Figure~\ref{fig:Steele-Path} depicts the dependence of $\mathcal{V}_{\rm He-\graphene}(x,y,z_0)$, i.e., the lateral dependence of the He--graphene minimum energy surface (values in Table~\ref{tab:VHeGraphene}) which allows for the calculation of $t_{\rm MP2}$ reported in \S{}\ref{subsec:bands}.

Additionally, we performed calculations for the energy for two He atoms adsorbed onto various hexagon centers.  After removing the baseline $2 \times \mathcal{V}_{\rm{He-\graphene}}$ terms, we find a remnant 
$\mathcal{V}_{\rm{He_\graphene}-\rm{He_\graphene}}(r)$
which remains strongly repulsive for nearest neighbors 
($r = 2.46$ \AA) and attractive for next-nearest neighbors and beyond ($r \ge 4.26$ \AA):
\begin{equation}
    V_{\rm MP2} = \SI{51.5}{\kelvin}, \quad V_{\rm MP2}^\prime = \SI{-1.97}{\kelvin}\, .
\label{eq:VBHMP2}
\end{equation}

\section{Discussion}
Our main result is the construction of a reliable and consistent description of
the effective two-dimensional adsorption problem of helium-4 on graphene in the
language of the hard-core Bose--Hubbard Model, \Eqref{eq:BHHamiltonian}. The
relevant hopping and interaction parameters computed via different techniques
are summarized in Table \ref{tab:BHresults}. The differences can be intuitively
understood by examining Fig.~\ref{fig:VHeGraphene}. For example, density
functional theory predicts a deeper (more attractive) He--graphene and He--He
potential (compared to the empirical, Lennard--Jones parametrized potential).
The resulting corrugation of the adsorption potential is enhanced (see
Table~\ref{tab:VHeGraphene}) leading to a suppressed hopping $t$, while the
increased two-body attraction between He atoms results in more spatially
localized wavefunctions (in the $xy$-plane) that yield a strongly reduced $V$, as the effects of the hard-core overlap are suppressed. 

On the other hand, the M\o{}ller--Plesset perturbative method gives the
strongest He--graphene interaction, leading to a smaller hopping $t$, while the
He--He interaction is close to the empirical one, and it gives similar values of $V$. Perhaps most importantly, the  quantum Monte Carlo and Hartree--Fock methods, both based on the empirical potentials, lead to similar results for all parameters. 

Overall, a remarkably consistent picture emerges. All of the above methods take into account the strong many-body He--He correlations on the scale of several \si{\angstrom},  comparable to the localization length of the one-particle wavefunctions in the lattice field of graphene. The simple one-particle Wannier description fails completely in this case, and thus the many-body techniques described in this work are essential to capture the self-consistent, interaction-driven adjustment of the one-particle orbitals, in turn leading to significant changes in the effective He--He interactions on the lattice scale.
 
To the best of our knowledge, this represents a unique case of a Bose--Hubbard model construction outside the usual examples that involve cold atom systems in optical lattice potentials.  Moreover, the Bose--Hubbard class of models that appear in cold atoms are much simpler to define and parametrize due to the diluteness of the atomic gases involved, which implies that the details of the atom--atom interactions on short scales are not important.  By contrast, for the case of helium on graphene, the details of the small distance He--He potential on the scale of the graphene lattice are extremely important, and consequently the effective Bose--Hubbard parameters are  very sensitive to the microscopic form of the potential employed. Due to the strong short-range repulsion, our model describes hard-core bosons ($U\approx\infty$), with finite nearest neighbor repulsion $V>0$ and much smaller next-nearest neighbor $V'<0$ attraction.  Our results place the first layer of He on graphene conclusively into the commensurate 1/3 filling insulating ground state on the triangular lattice formed by the centers of graphene hexagons.
 
Armed with the above realization,  we envisage avenues of research that involve effective Bose--Hubbard Hamiltonians of atoms on 2D materials with different lattice parameters.  Numerous 2D materials exist, and in addition,  their parameters can be affected by external knobs such as strain, doping, etc. These factors also affect the strength of the atom--material potential (which is of van der Waals origin).  The ultimate advantage of having a reliable effective Bose--Hubbard description is that it allows studies of strongly correlated phases, such as supersolids, correlated insulators and superfluids,  as well as the phase transitions between them. Thus  Bose--Hubbard model construction can be viewed as a project of designing low-dimensional physical systems with given correlated ground state properties, e.g., superfluids in a regime (density, temperature, size) more aligned with conventional solid state physics.   

Finally, we mention that the route towards such designer Hamiltonians is more complex than the usual ``band-structure engineering" which relies on the numerical construction, for example, of maximally localized single-particle Wannier states. The accurate determination of interaction parameters as described in this work adds additional computational complexity due to the need to carefully incorporate the effects of many-body interactions --- it can range from a modest one for the Hartree--Fock implementation to a large-scale use of computational resources for quantum Monte Carlo or ab initio methods.

\acknowledgments

We thank our late colleague, Dr.~Darren Hitt, former director of the VT Space Grant Consortium for his leadership, encouragement, and vision to expand the scope of space grant activities in Vermont, all essential to forming our interdisciplinary collaboration.  This work was supported, in part, under NASA grant number 80NSSC19M0143.  N.S.~Nichols acknowledges partial support from the National Science Foundation (NSF) under award No.~DMR-1808440.  Computational resources were provided by the NASA High-End Computing (HEC) Program through the NASA Advanced Supercomputing (NAS) Division at Ames Research Center. 

\appendix

\section{Dimensional Reduction of the Adsorption Layer to 2D}
\label{app:3Dto2D}

The He--graphene potential in Eq.~\eqref{eq:HeGraphene} can be written as
\be
\mathcal{V}_{\rm He-\graphene}(\vec{r}) = \mathcal{V}_0(z) + \mathcal{V}_{\perp}(\vrg,z),
\label{eEmp1_01}
\ee
where $\mathcal{V}_0(z)$ and $\mathcal{V}_{\perp}(\vrg,z)$ are the
$\vec{g}=0$ and $\vec{g}\neq 0$ terms, respectively. 
A justification for such a splitting was presented in \S{}\ref{sec:monolayer}. 
We seek the single-particle minimizer of the Hamiltonian in Eq.~\eqref{eq:Ham}, \emph{i.e.}~the solution of:
\be
-\frac{\hbar^2}{2m} \nabla^2 \Psi(\vec{r})  + 
 \mathcal{V}_{\rm He-\graphene}(\vec{r}) \Psi(\vec{r}) = E \Psi(\vec{r}),
\label{eEmp1_02}
\ee
as a (truncated) expansion over eigenfunctions $\{\phi_n(z)\}$ of the 1D potential $\mathcal{V}_0(z)$:
\bsube
\be
\Psi(\vec{r}) \equiv \Psi(\vrg,z) = \sum_{n} \chi_n(\vrg) \phi_n(z)\,;
\label{eEmp1_03a}
\ee
where $\phi_n$ satisfy 
\be
-\frac{\hbar^2}{2m}\phi_n'' + \mathcal{V}_0 \phi_n = \epsilon_n \phi_n, 
\qquad 
\langle \phi_n|\phi_n \rangle = 1\,,
\label{eEmp1_03b}
\ee
\label{eEmp1_03}
\esube
and $\langle \ldots \rangle$ stands for integration over $z$. 
Substituting \Eqref{eEmp1_03a} into \Eqref{eEmp1_02}, multiplying by $\langle \phi_n|$ and integrating over $z$, one obtains a system of coupled equations for $\chi_n\qty(\vrg)$. 
Such a system can, in principle, 
be solved by the same numerical method as described
in \S{}\ref{subsec:bands}. To focus on the conceptual consequences of the
$z$-spread of the wavefunction rather than on finer details,
we proceed by truncating the expansion in \Eqref{eEmp1_03a} at lowest order:
\bsube
\be
\Psi\qty(\vrg,z) \approx \chi_0(\vrg) \phi_0(z)\,.
\label{eEmp1_04a}
\ee
The quantity $\chi_0\qty(\vrg)$ plays the role of an effective ``reduced"  2D wavefunction and satisfies the Schr\"odinger equation:
\begin{align}
    \label{eEmp1_04b}
    -\frac{\hbar^2}{2m}\nabla_{\vrg}^2\chi_0 & + \tilVHeG \;\chi_0 = \tilde{E} \chi_0\, , \\ 
    \tilVHeG\qty(\vrg) &\equiv \expval{\mathcal{V}_0(z) + \mathcal{V}_{\perp}(\vrg,z)}{\phi_0}\, ,
\label{eEmp1_04c}
\end{align}
\label{eEmp1_04}
\esube
where $\nabla_{\vrg}^2$ is the Laplacian in $\vrg$.
We have absorbed  a constant $\expval{\mathcal{V}_0(z)}{\phi_0}$ into both sides
of \eqref{eEmp1_04b} to obtain correspondence with the quantum
Monte Carlo $z$-averaging results described in \S~\ref{subsec:qmc}.
Equation \eqref{eEmp1_04b} 
represents the 2D reduction of the 3D one-particle model,
which is solved in \S{}\ref{subsec:bands}.

We note that the particle density corresponding to the approximation \eqref{eEmp1_04a} is:
\be
\rho(\vrg,z) = |\psi_0(\vrg)|^2 \rho(z), 
\qquad 
\rho(z) \equiv |\phi_0(z)|^2\,.
\label{eEmpadd1_01}
\ee
Substituting this into \eqref{eq:VHeGrapheneAverage}, one finds that
the expression there coincides with $\tilVHeG$ in \eqref{eEmp1_04c}.
The approximate particle density in the $z$-direction,  $\rho(z)$, can be found by solving Eq.~\eqref{eEmp1_03b} with $n=0$ by, \emph{e.g.}, the shooting method. 
The result is shown in Fig.~\ref{fig:ShootingMethod}.

The spread of the single-particle density $\rho(z)$ in the $z$-direction also affects the computation of the effective nearest neighbor $V$ in the Bose-Hubbard model. This spread leads to nearest-neighbor $^4$He atoms having a distribution of $z$-values relative to one another \cite{Vidali:1980pg} in adjacent graphene hexagons:
\be
\gamma(\delta) = \int \rho(z) \rho(z+\delta) \,dz\,.
\label{eEmp1_05t}
\ee
Here we have assumed, in agreement with the finding by QMC simulations, 
that there is no correlation
between $z$-values of the centers of nearest-neighbor $^4$He atoms. 
This quantity 
 can then be used to estimate the effect of the relative $z$-spread $\delta$ 
of nearest-neighbor $^4$He atoms on their interaction via:
\be
\widetilde{\mathcal{V}}_{\rm He-He}(\vrg) = \int \Vhehe\big(\sqrt{|\vrg|^2 + \delta^2}\big)\,
\gamma(\delta) \, d\delta\,.
\label{eEmp1_06}
\ee
It leads to some ``softening" of the He--He interaction potential.
In \S{}~\ref{subsec:HF} we show how much this softening affects $V$. 
Arguably, to more properly account for the spread in the $z$-direction, instead of using the $\rho(z)$ defined in \eqref{eEmpadd1_01} one would need to use $\rho(z)$ obtained by Monte Carlo simulations (\S{}\ref{subsec:qmc}).  The results of using both approaches are compared in \S{}\ref{subsec:HF}.

\section{Simulation Details and Scaling}
\label{app:QMCscaling}

In this appendix, we provide details on the quantum Monte Carlo method used in \S~\ref{subsec:qmc}.  Access to the employed software can be obtained via Ref.~\cite{delmaestroCode}.

\subsection{Algorithmic Convergence}
By choosing a suitably small imaginary time step $\tau$ (polynomial scaling) and long enough imaginary time projection length $\beta$ (exponential dependence) it is possible to ensure that any systematic errors inherent in the choice of an approximate propagator $\rho_\tau$ \cite{Chin:1997cu,Jang:2001cl} are made smaller than any statistical errors in our ground state quantum Monte Carlo scheme.   At the additional computational expense of requiring a potentially larger value of $\beta$ to obtain convergence, we have chosen to employ the constant trial wavefunction $\Psi_T(\vec{R})=1$ to prevent the breaking of translational symmetry of the adsorbed phase in order to explore the effects of particle tunneling. 

\begin{figure}[ht]
\begin{center}
    \includegraphics[width=\columnwidth]{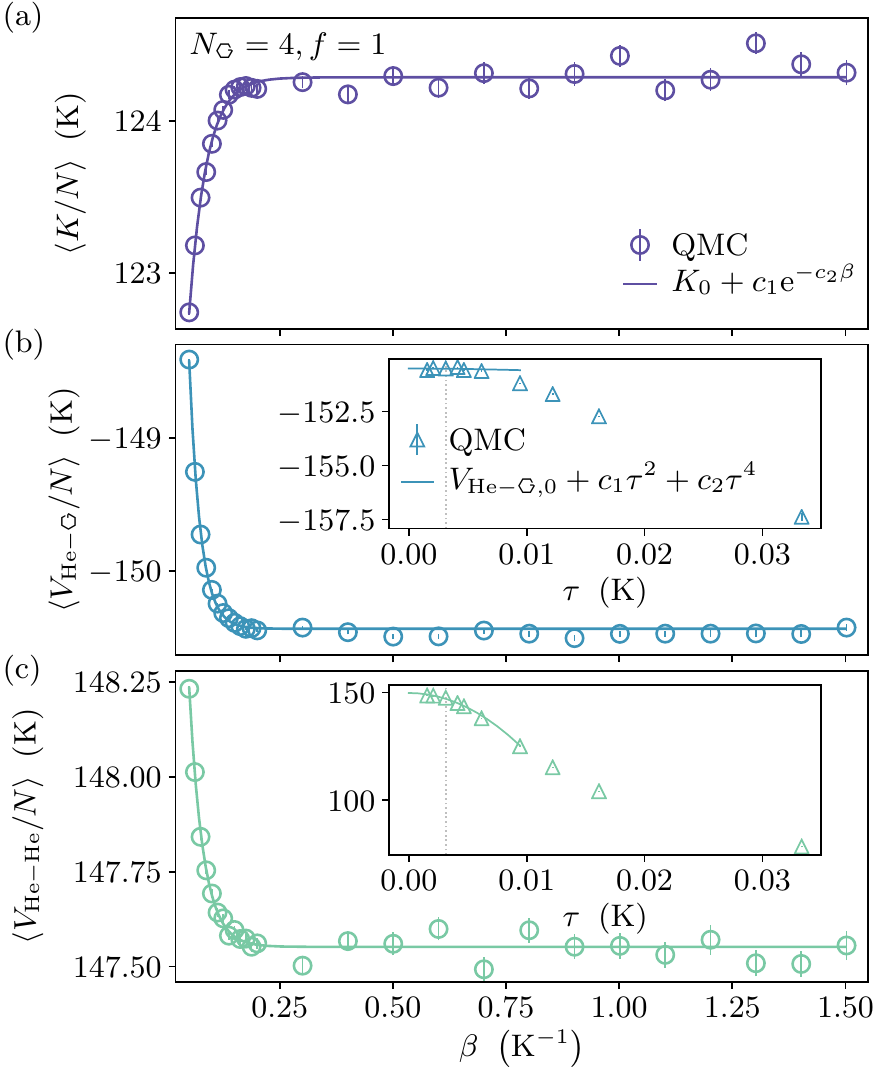}
\end{center}
\caption{Convergence of the kinetic (a), adsorption (b), and interaction potential (c) energy per particle with the imaginary time projection length $\beta$ and imaginary time step $\tau$ (insets) for a system with $N_\graphene = 4$ adsorption sites at unit filling.  All quantities measured via quantum Monte Carlo (symbols + errorbars) are observed to converge to their ground state value with the expected exponential dependence on $\beta$ or polynomial dependence on $\tau$ as quantified via the indicated fits (lines). $\beta$ scaling was performed with $\tau = \SI{1/319}{\kelvin}$ (vertical dashed line in insets) while $\tau$ scaling had $\beta \simeq \SI{0.5}{\kelvin^{-1}}$. A subscript $0$ indicates the ground state value.} 
\label{fig:betatauScaling}
\end{figure}

Figure~\ref{fig:betatauScaling} shows the convergence of the energy as a function of $\tau$ and $\beta$.  
%
%
We choose $\tau \simeq \SI{0.00313}{\kelvin}$ corresponding to $319$ discrete imaginary time steps and $\beta \simeq \SIrange{0.5}{1.0}{\kelvin^{-1}}$ for all simulations presented in this work.  

\subsection{Finite Size Scaling}
\begin{figure}[ht]
\begin{center}
    \includegraphics[width=\columnwidth]{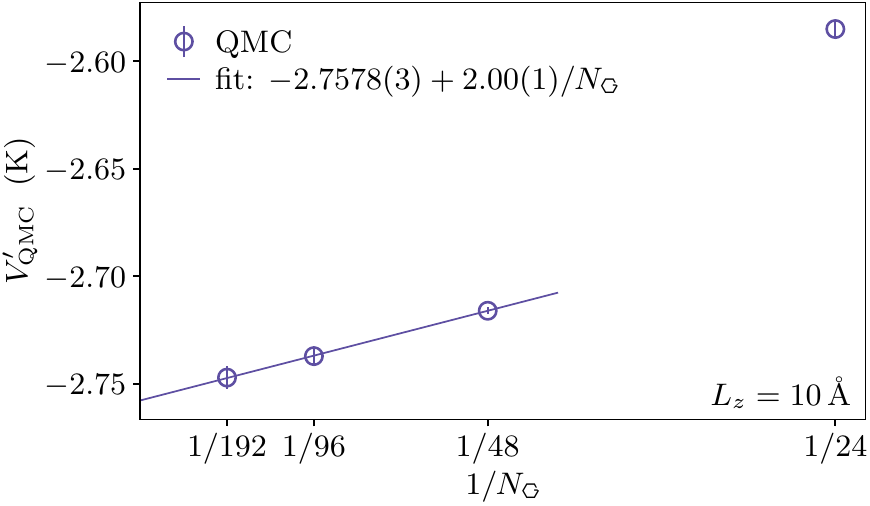}
\end{center}
\caption{The finite size scaling of the next nearest neighbor interaction
$V^\prime$ extracted from quantum Monte Carlo as a function of the number of
graphene hexagon adsorption sites.  The solid line shows a linear fit to the
data for $N_\graphene \ge 48$ which allows extrapolation to the thermodynamic limit.}
\label{fig:Vprimefss}
\end{figure}

While the spatial extent of the simulation cell in the $x$ and $y$ directions had a minimal effect on most observables (see discussion and results in \S~\ref{subsec:qmc}) there was some observed dependence of $V^\prime$ on $N_\graphene$ as shown in Fig.~\ref{fig:Vprimefss}.
%
%
This is likely due to the finite size configuration at $f=1/3$ tunneling between the three equivalent configurations in the commensurate cell which would be suppressed in the thermodynamic limit.  We have employed a linear extrapolation to $N_\graphene = \infty$ to obtain the reported result for $V^\prime$. 

\subsection{Stabilizing a Single Adsorbed Layer at $f=1$}
\label{subsec:singleLayer}

\begin{figure}[t]
\begin{center}
    \includegraphics[width=\columnwidth]{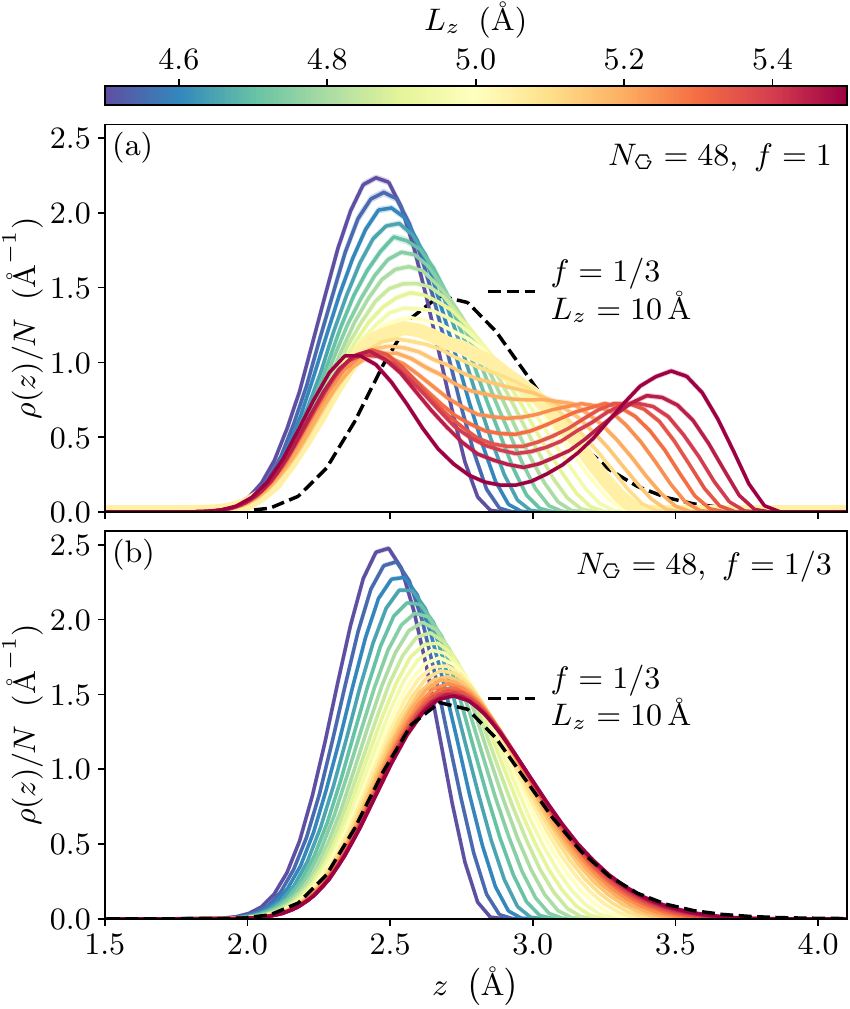}
\end{center}
\caption{The density profile (per particle) of the adsorbed layer(s) for
different vertical box sizes $L_z$ enforced through the potential in
\Eqref{eq:Vwall} for helium above a graphene sheet with $N_\graphene = 48$
adsorption sites such that $L_x \times L_y = \SI{14.75707}{\angstrom}\times
\SI{17.04}{\angstrom}$. Panels correspond to (a) filling $f=1$ and (b) $f=1/3$ where statistical uncertainties are indicated by the shaded envelope. The thicker curve in (a) for $L_z=\SI{5.05}{\angstrom}$ was determined to be the optimal value (see text). The dashed line indicates the density profile for a ``bulk'' cell with $L_z = \SI{10}{\angstrom}$ at $f=1/3$ that is used for comparison. The number of $^4$He atoms in the simulation can be determined from $N= f N_\graphene$.}
\label{fig:layering}
\end{figure}

\begin{figure}[t]
\begin{center}
    \includegraphics[width=\columnwidth]{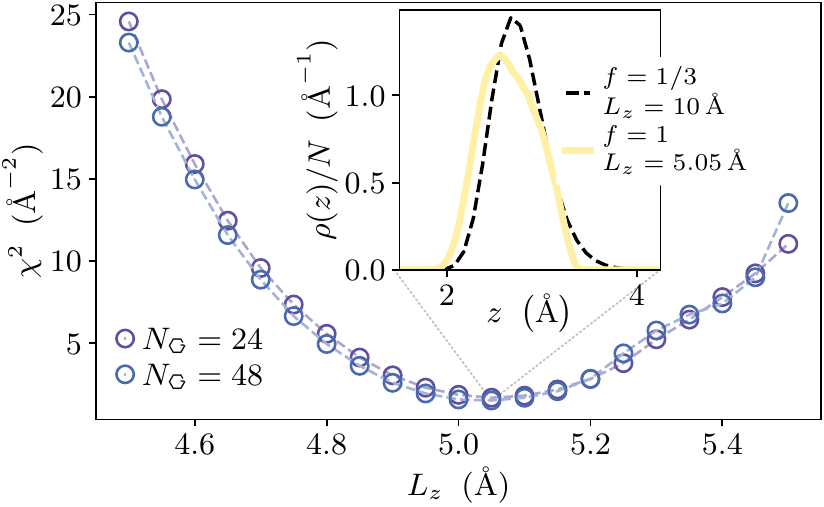}
\end{center}
\caption{The squared deviation between solid and dashed curves in 
    Fig.~\ref{fig:layering}(a) as a function the box size in the $z$-direction as
quantified in \Eqref{eq:chi2}.  The minimum at $L_z =
\SI{5.05}{\angstrom}$ is independent of the size of the
graphene sheet, where data for $N_\graphene = 24$ and $48$ are shown. The inset shows a comparison of the density profiles in the $z$-direction for this value at $N_\graphene = 48$. The dashed line is a guide to the eye.}
\label{fig:chi2}
\end{figure}

In order to answer the questions posed in \S{}~\ref{subsubsec:manybody}, we have performed QMC simulations at filling fractions:
$f=1/3,1$ for $L_z \in [4.5,5.5]$ and $N_\graphene = 24,48,96$ (parallelization of analysis was accelerated using GNU Parallel \cite{tange_2020_4045386}).
Finite size effects in $N_\graphene$ were negligible for the density profiles in the $z$-direction, and we show simulation results for $N_\graphene = 48$ in Fig.~\ref{fig:layering}.  Here panels correspond to different filling fractions and colors to different values of $L_z$.
%
%
In panel (a) at unit filling ($f=1$) we observe that the density $\rho(z)$
smoothly evolves as a function of $L_z$ from one that contains a single
well-defined layer for small box sizes ($L_z \lesssim \SI{5}{\angstrom}$), to a
profile with two peaks in the density for $L_z \gtrsim \SI{5}{\angstrom}$.  In
order to quantify these two regimes and determine at which value of $L_z$ we
should analyze the system, we performed additional simulations at $f=1/3$
(Fig.~\ref{fig:layering}(b)) where we observe less drastic effects of the
confinement.  At this lower filling, results are clearly approaching the bulk
case for $f=1/3$ with $L_z=\SI{10}{\angstrom}$ beyond $L_z \gtrsim
\SI{5.5}{\angstrom}$ as indicated by the dashed line.  This data can then be
exploited by searching for the value of $L_z$ at unit filling that produces a
density profile most similar to that of the bulk monolayer at $f=1/3$ within
the approximation that interactions in the plane should not seriously affect
the $z$-spread of the wavefunction. To proceed, we search for a minimum in the squared deviation of densities:
\begin{equation}
    \chi^2(L_z) \equiv \sum_i \qty[\eval{\frac{\rho(z_i)}{N_\graphene}}_{\substack{L_z \\ f=1}}-\eval{\frac{\rho(z_i)}{N_\graphene/3}}_{\substack{Lz=\SI{10}{\angstrom} \\ f=1/3}}]^2 \label{eq:chi2}
\end{equation}
where $i$ runs over all spatial positions in $z$ where density data has been obtained. The results of this procedure are shown in Fig.~\ref{fig:chi2} 
%
%
and indicate a quadratic dependence on $L_z$ with the minimum occurring at $L_z = \SI{5.05}{\angstrom}$. At this value of $L_z$, the inset shows a comparison of the two density profiles from Fig.~\ref{fig:layering}.  Finite size effects in $N_\graphene$ were not found to alter the optimal value of $L_z$. 

\begin{figure}[h]
\begin{center}
    \includegraphics[width=\columnwidth]{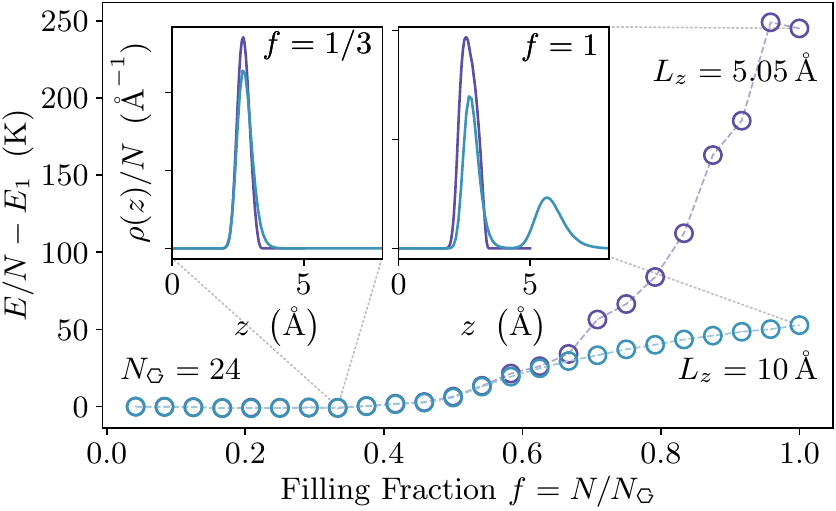}
\end{center}
\caption{Equation of state (energy per particle as a function of filling fraction) for $N_\graphene = 24$ adsorption sites for two box sizes with $L_z = \SIlist{5.05;10.0}{\angstrom}$. The curves have been shifted by the energy for a single particle $N=1$ corresponding to a filling fraction-independent value of $\sim \SI{6}{K}$ due to the presence of $\mathcal{V}_{\rm wall}$.  The insets show the density of particles along the $z$-direction at filling fractions $f=1/3$ and $1$. For unit filling, the cell with $L_z=\SI{10}{\angstrom}$ can accommodate a second layer.}
\label{fig:Eos}
\end{figure}

Recall that the goal of this procedure was to stabilize a single monolayer at
filling fraction $f=1$ in order to determine the effects of nearest and
next-nearest neighbor interactions between $^4$He atoms without substantially distorting the physics of the adsorbed phase.  As an additional check, we have computed the equation of state at $L_z=\SI{5.05}{\angstrom}$ and compared it with that determined for the unrestricted bulk cell with $L_z = \SI{10}{\angstrom}$ for a system with $N_\graphene = 24$ adsorption sites.  The results, shown in Fig.~\ref{fig:Eos}, demonstrate that the additional confinement potential in \Eqref{eq:Vwall} does not alter the ground state properties of the adsorbed monolayer for $f \lesssim 0.6$.
%
%
The insets show that the monolayer profile remains mostly unchanged for filling
fraction $f=1/3$. At unit filling with $f=1$, while the confined box with $L_z
= \SI{5.05}{\angstrom}$ still exhibits only a single layer, the unbounded cell
can now accommodate an energetically favorable second layer.


\section{Quasiclassical approximation}
\label{app:WKB}

%
\begin{figure}
\includegraphics[width=\columnwidth]{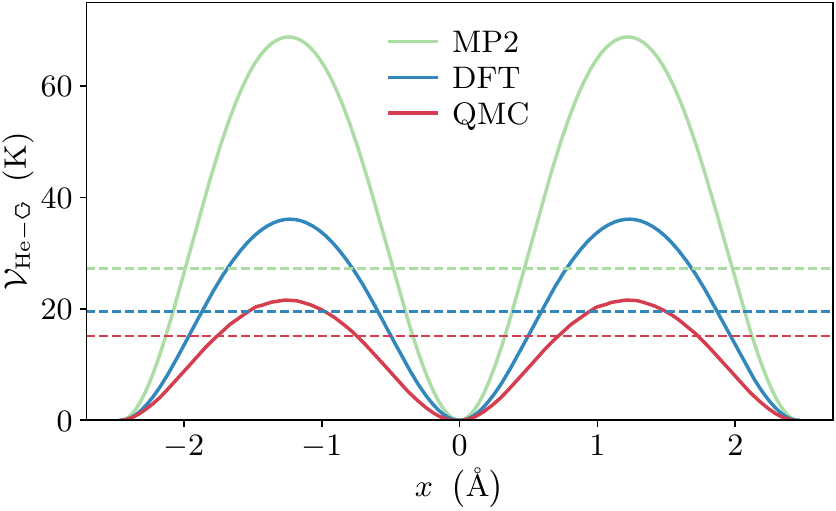}
\caption{A one dimensional cut through $\Vext$ for a $^{4}$He
    atom above graphene corresponding to the spatial path through the saddle
    point of the full 2D potential shown in Fig.~\ref{fig:Steele-Path}.  The
    energy barriers that inhibit tunneling between different graphene
    adsorption sites originate from three different methods: M\o{}ller--Plesset
    perturbation theory (MP2, green), density functional theory (DFT, blue), and quantum Monte Carlo (QMC, red) as described in the text, where the curves have been shifted with respect to the bottom of the potential well. The classical turning points $x_{c_{1,2}}$ for each potential barrier appear at the intersection of the dashed 
($E_{0}=\hbar\omega_{0}/2$) and solid lines.
 }
\label{fig:Tunneling}
\end{figure}
%

The one-dimensional (1D) WKB approach is a semiclassical approach which
can portray tunnel splitting and is well defined in 1D double-well
or periodic potentials.
In order to obtain an intuitive and simple estimate of the hopping $t$ we will
apply the 1D approach along the path  passing through the saddle point in our 2D potential, Fig.~\ref{fig:Steele-Path}.
This path  connects two adjacent minima (Fig.~\ref{fig:Tunneling}) and leads to the largest hopping amplitude.

In this quasi 1D limit the energy dispersion along any of the three triangular lattice  directions has the form: 
$\varepsilon\left(k\right)-\varepsilon_{0} = -2t\cos\left(k a\right)$.
The quasiclassical expression for the hopping is known \cite{LL9,Holstein:1988} to be:
\begin{equation}
t=\frac{\hbar\omega_0}{2\pi}\exp\left(-\frac{1}{\hbar}\int_{x_{c_1}}^{x_{c_2}}\sqrt{2m\left(\Vext\left(x\right)-\hbar\omega_{0}/2\right)}dx\right), 
\label{eq:WKB}
\end{equation}
where the classical turning points $x_{c_{1,2}}$ satisfy $\Vext\left(x_{c_{1,2}}\right)=\hbar\omega_{0}/2$
and the integral is over the barrier interior.  Here $\omega_{0}$
is the frequency of small amplitude oscillations in the wells which are fitted
to parabolic (harmonic oscillator) form. This expression is valid as long as
the potential barrier is high enough and the exponential tunneling factor is
small, which is only approximately satisfied in our system. Overall the hopping is a product of the tunneling factor and the attempt frequency $\omega_0$, leading to a finite number expected to provide a good numerical estimate.  The hopping parameter derived from He--graphene interactions obtained by various methods (Fig.~\ref{fig:Tunneling}) are shown in Table \ref{tab:WKB}.  It is clear that the results from this simple approximation provide quite reasonable estimates as they are comparable to the numbers and tendencies from the full 2D calculations whose results are displayed in Table \ref{tab:BHresults}.
\begin{table}
    \renewcommand{\arraystretch}{1.5}
    \setlength\tabcolsep{12pt}
    \begin{tabular}{@{}lllll@{}} 
        \toprule
        Method & ${\hbar\omega_{0}}/{2} \, (\si{\kelvin})$ & $t \,
        (\si{\kelvin})$ & $t/V$ \\ 
        \midrule
        QMC & 15.1 & 2.22 & 0.041 \\
        DFT  & 19.6 & 1.25 &  0.058 \\
        MP2  & 27.3 & 0.43 &  0.008 \\
        \bottomrule
    \end{tabular}
    \caption{\label{tab:WKB} 
    Ground state energy in the potential wells in the harmonic approximation ($\hbar\omega_{0}/2$) and
   quasi-classical   hopping parameters $t$, Eq.~(\ref{eq:WKB}), derived for He--graphene interactions obtained by various methods, Fig.~\ref{fig:Tunneling}.}
\end{table}

\nocite{apsrev42Control}
\bibliographystyle{apsrev4-2}
\bibliography{refs}

\end{document}